\documentclass{optica-article}

\journal{opticajournal} 

\articletype{Research Article}

\usepackage{lineno}
\usepackage{hyperref}

\begin{document}

\title{Can We Conjugate the Orbital Angular Momentum of a Single Photon with Nonlinear Optics? A Theoretical Analysis}


\author{Tobey J.B. Luck,\authormark{*1} Aarón A. Aguilar-Cardoso,\authormark{1} Cheng Li,\authormark{1} Manuel F. Ferrer-Garcia,\authormark{1} Jeremy Upham, \authormark{1} Jeff S. Lundeen\authormark{1,2}, Robert W. Boyd\authormark{1,3}}

\address{\authormark{1} Nexus for Quantum Technologies, Department of Physics, University of Ottawa, Ottawa, K1N 6N5, ON, Canada\\
\authormark{2} National Research Council of Canada, 100 Sussex Drive, Ottawa, Ontario K1A 0R6, Canada.\\
\authormark{3} Institute of Optics, University of Rochester, Rochester, 14627, NY, USA.}

\email{\authormark{*}tluck074@uottawa.ca} 


\begin{abstract*} 
Stimulated parametric down-conversion (StimPDC) produces an idler beam that exhibits a phase conjugate of the spatial structure of the classical seed beam. Extending this effect to the single-photon regime can, in principle, enable the control of the orbital angular momentum of single photons. However, its experimental implementation remains challenging due to the lower photon flux of realistic single-photon sources compared to a classical pump beam, which causes the StimPDC process to occur less frequently compared to its spontaneous counterpart. To overcome this, we propose a cascaded parametric down-conversion scheme, in which the signal photon from a spontaneously down-converted pair seeds the StimPDC process, while the idler photon heralds the seed’s presence. This configuration uses coincidence detection to effectively post-select for stimulated events among the background of spontaneous contributions. We show that phase conjugation at the single-photon level manifests as a bias towards equal orbital angular momentum between the heralding idler and the stimulated idler. Our results offer a practical route to manipulate single-photon correlations in the orbital angular momentum degree of freedom and could have implications for applications such as free-space quantum communication and quantum-enhanced sensing.
\end{abstract*}

\section{Introduction}
The family of optical beams carrying orbital angular momentum (OAM) of light, characterized by a discretized azimuthal phase dependence $\exp{\{i \ell \phi\}}$, has emerged as a pivotal resource in modern optics. Photons carrying OAM enable high-dimensional encoding of information, making them invaluable for applications ranging from quantum communication, cryptography, and sensing technologies \cite{Cozzolino:2019, Vallone:2014, Mirhosseini:2015, Maga:2014}. The unbounded nature of the OAM quantum number $\ell$ further allows for dense multiplexing in classical and quantum channels, enabling high-capacity data transmission and enhanced noise resistance \cite{Yan:2014, Ecker:2019}. However, manipulation of OAM states at the single-photon level, such as phase conjugation, remains experimentally challenging, limiting their potential in practical quantum information technologies.

One transformation of particular relevance to OAM-based systems is the phase conjugation of the azimuthal structure of light, in which a light beam carrying topological charge $\ell$ is converted into one with charge $-\ell$. This process reverses the helical phase front of the input beam, effectively flipping the direction of OAM while preserving the radial structure. We note that one can extend this concept to reverse spatial distortions or wavefront curvature in arbitrary spatial modes of light \cite{Bogdanov:1996, Liu:2017, Yu:2019}. Here, OAM phase conjugation refers specifically to the inversion of the azimuthal phase dependence $\exp\{i\ell\phi\}$, which can also apply to superpositions of OAM, each $\ell$ maps to $-\ell$, and the coefficients are complex-conjugated. This effect has been leveraged to design OAM multiplexers, quantum state synthesizers, and entanglement switchers, underscoring its versatility in both classical and quantum regimes \cite{Yan:2014, Simon:2015, Nagali:2009}. Phase conjugation of OAM can be implemented through processes such as difference-frequency generation (DFG) \cite{Fewings:00}, where angular momentum conservation enforces conjugation between input and output modes. While these effects have been demonstrated in classical regimes using coherent beams, their behaviour in the single-photon regime remains largely underexplored. Phase conjugation on the single-photon level is of particular interest due to its potential for practical quantum applications. For instance, the deterministic unitary induced on the OAM labels, which is the mode inversion $\ell \to -\ell$, can be used as a building block for quantum gates in OAM encodings (distinct from full complex conjugation, which is non-unitary) \cite{Gasparoni:2004}. In quantum communication, single-photon phase conjugation could be used to extend various protocols into the single-photon regime. One such protocol is the mitigation of atmospheric turbulence-induced phase distortions in free-space OAM channels by mixing the phase conjugate of a probe beam (which characterizes the turbulence of the channel) with information to be sent, causing the phase distortion to be reversed upon propagation back through the turbulent media \cite{Zhou:23}. More broadly, access to quantum-compatible phase conjugation enables new strategies for high-fidelity information transfer, such as suppression of Kerr nonlinearities in optical fiber communications and adaptive optics within quantum networks \cite{Zhou:21, Liu:2013, Leonhard:2018}.

Single-photon-stimulated parametric down-conversion (SPS-PDC), which refers to the single-photon-seeded counterpart of classically seeded StimPDC, provides a route to OAM mode phase conjugation at the single-photon level. Here, SPS-PDC denotes a $\chi^{(2)}$ interaction pumped by a classical field and stimulated by an input single photon. In conventional classically seeded StimPDC, injecting a bright seed into a pumped crystal produces down-converted light in which the idler carries the phase conjugated spatial mode of the seed \cite{Riberiro:1999, Xu:23}. In the single-photon-seeded regime, the stimulated contribution competes with a substantial spontaneous background, unlike the seed-dominated output of StimPDC \cite{Permaul:2025}. Both StimPDC and SPS-PDC offer several practical advantages over other methods of phase conjugation. Unlike digital approaches based on spatial light modulators, they are all-optical $\chi^{(2)}$ interactions whose response is set by the pump and phase matching bandwidths, allowing sub-nanosecond pump-limited operation without wavefront measurement \cite{Wang:2015, Cerullo:2003}. Compared with $\chi^{(3)}$ four-wave mixing, $\chi^{(2)}$ implementations in bulk crystals avoid resonant media and associated complexity while maintaining high spatial-mode fidelity \cite{Turnbull:2013, Qiu:2023}. Moreover, energy conservation $(\omega_i=\omega_p-\omega_s)$ yields simultaneous phase conjugation and frequency conversion, enabling translation between spectral bands without mode-locked sources \cite{Tsang:2006}. To experimentally observe azimuthal phase conjugation in this quantum regime, it is essential to distinguish relatively rare stimulated events from a much more prevalent spontaneous background, which is a challenge arising from the inherently low intensity of the seed relative to the pump.

In this work, we analyze the SPS-PDC process and propose a cascaded implementation. The configuration, which is denoted by cascaded parametric down-conversion (CPDC), first generates a pair of down-converted photons via spontaneous parametric down-conversion (SPDC); one photon seeds an SPS-PDC process in a second nonlinear crystal, while the presence of the seed photon is heralded by detection of its idler partner. The SPDC process thus serves as both a single-photon source and a source of timing information, which allows us to measure the OAM spectrum of the SPS-PDC process conditioned on the presence of a seeding photon. This paper is organized as follows. In Section 2, we derive the mean number of idler photons in the SPS-PDC process in a specific Laguerre-Gaussian (LG) mode and find that the phase conjugation effect corresponds to higher mean photon numbers when the OAM of the idler and seed photons are conjugate. We then show that without heralding, the stimulated phase conjugation of the seed is overwhelmed by spontaneous emission when the seed flux is significantly lower than the pump flux, rendering the phase conjugation effect difficult to observe directly. In Section 3, we show that the CPDC approach effectively filters out unseeded spontaneous events by heralding the presence of a seed, thereby post-selecting for the stimulated events in which the phase conjugation is observed. We develop a theoretical framework for identifying signatures of OAM phase conjugation in the quantum regime and assess the conditions under which the effect can be unambiguously observed. Additionally, we quantify the visibility of the phase conjugation effect in both configurations and demonstrate how the heralding of the seed allows for the excess spontaneous events to be neglected, thus greatly increasing the visibility of the desired phase conjugation effect. In Section 4, we present our conclusions.

\section{Single-photon-stimulated parametric down-conversion without heralding}
\begin{figure}
    \centering
    \includegraphics[width=0.8\linewidth]{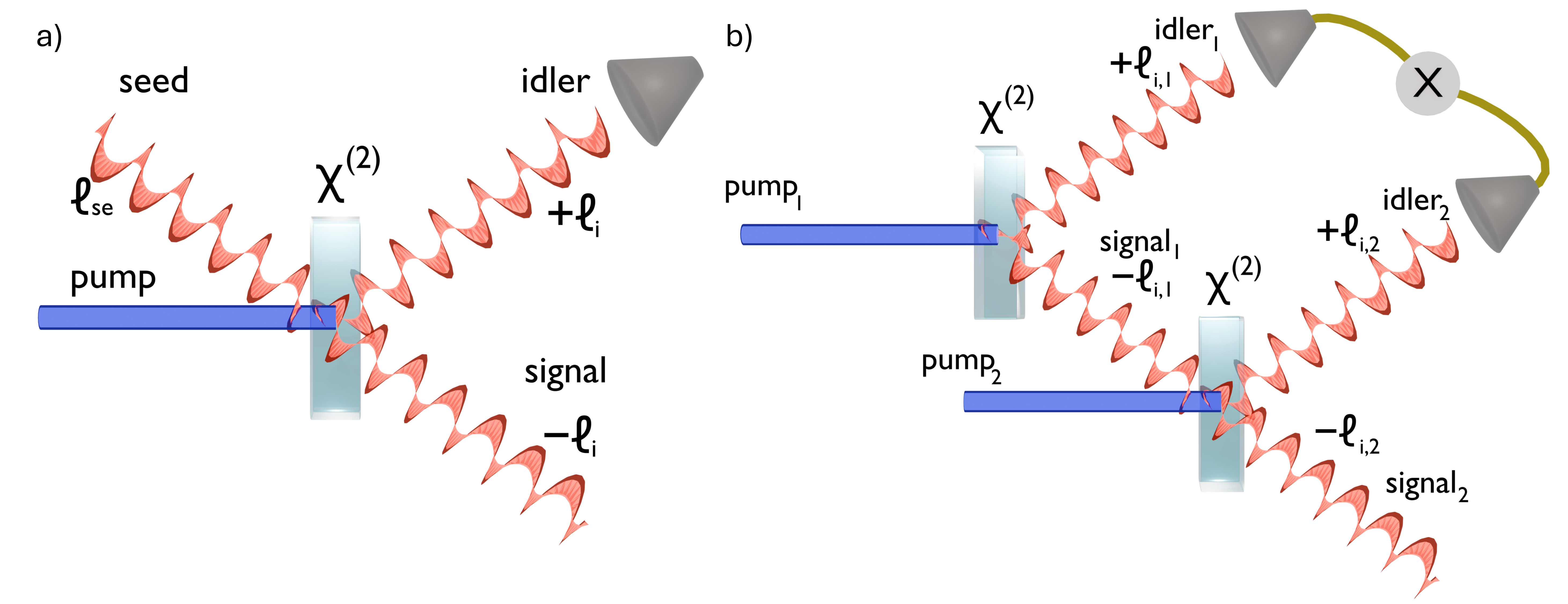}
    \caption{Schematic of the a) SPS-PDC process in which the mean number of photons in a specified LG mode of the idler field is measured; b) CPDC process in which a single photon undergoes azimuthal phase conjugation. A seed photon (signal$_1$) generated via SPDC seeds the nonlinear process in the second crystal, while its partner photon (idler$_1$) serves as a herald. Coincidences are measured between the heralding photon (idler$_1$) and the idler photon generated by the SPS-PDC process (idler$_2$), allowing the OAM spectrum of the SPS-PDC process to be measured conditioned on the presence of the seed photon.}
    \label{fig:StimPDC}
\end{figure}
\subsection{Mean idler photon number distribution in Laguerre-Gaussian modes}
\begin{figure}
    \centering
    \includegraphics[width=0.9\linewidth]{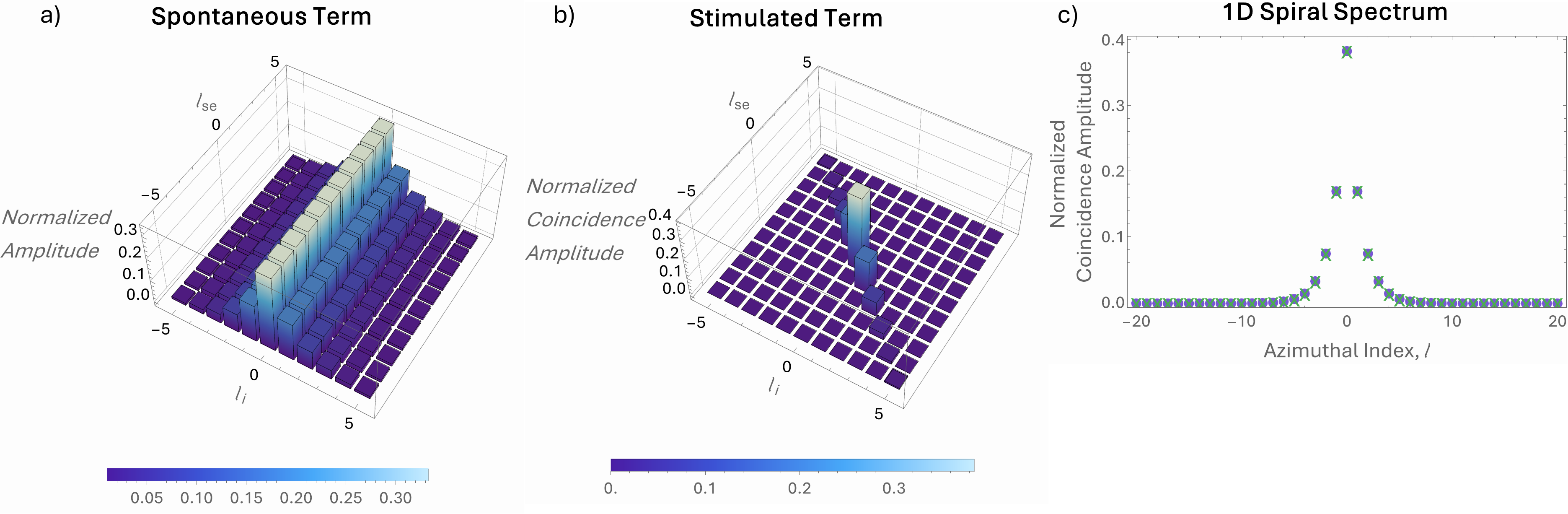}
    \caption{Signature of OAM phase conjugation in SPS-PDC. a) The spontaneous term: the idler distribution is independent of the seed’s OAM. b) The stimulated term only contributes along the anti-diagonal $\ell_i = -\ell_{se}$, the signature of OAM phase conjugation. c) 1D spiral spectrum: the anti-diagonal slice (purple dots) agrees with the known SPDC spiral spectrum (green crosses \cite{Miatto:2011}). All plots are normalized over all possible azimuthal indices and $p_i = p_{se} = 0$.}
    \label{fig:spiralbandwidth}
\end{figure}

In SPDC, the OAM of the down-converted signal and idler photons is related to the OAM of the pump by the conservation of OAM, $\ell_p = \ell_s + \ell_i$, where $\ell_p, \ell_s, \ell_i$ denote the OAM of the pump, signal, and idler, respectively \cite{Mair:2001}. When the OAM of the pump is zero, the signal and idler photons will have conjugate OAM, $\ell_s = -\ell_i$, or equal in magnitude and opposite in sign. The down-converted photons generated in conjugate OAM states following the distribution known as the spiral spectrum, denoted $\mathcal{M}^{\ell_i,\ell_s}_{p_i,p_s}$, which relates the probability of generation to azimuthal indices $\ell_i$, $\ell_s$ and radial indices $p_i$, $p_s$ \cite{Torres:2003, Miatto:2011}. In SPS-PDC, the conservation of OAM between the pump, signal, and idler also applies \cite{Caetano:2002}. However, in this section, we show how the OAM spectrum of down-converted photons does not directly follow the spiral spectrum. This is done by deriving the mean photon number of idler photons in mode $\tilde{LG}^{\ell_i}_{p_i}$ in the SPS-PDC process where we find that the mean photon number is the sum of a spontaneous and stimulated term, where $\tilde{LG}^{\ell_i}_{p_i}$ represents a LG mode in \textit{k-space} (in what follows, `$\sim$' denotes that the quantity is defined in $k$-space) with azimuthal index $\ell_i$ and radial index $p_i$. The normalized LG mode functions in $k$-space are given by
\begin{equation}
    \label{LGMode}
    \tilde{LG}_p^{\ell}(\rho, \varphi) = \sqrt{\frac{w^2 p!}{2\pi (p + |\ell|)!}} \left( \frac{\rho w}{\sqrt{2}}\right)^{|\ell|} e^{-w^2\rho^2/4} (-1)^p L_p^{|\ell|} \left( \frac{\rho^2 w^2}{2}\right) e^{\left[ i\ell\left(\varphi + \frac{\pi}{2}\right)\right]},
\end{equation}
where $L_p^{|\ell|}(\cdot)$ is the associated Laguerre polynomial. The SPS-PDC process is depicted schematically in Fig.~(\!\hyperref[fig:StimPDC]{~\ref{fig:StimPDC}a}), where a detector measuring mode $\tilde{LG}^{\ell_i}_{p_i}$ measures the output of the idler in the SPS-PDC process. Here, the SPS-PDC process is seeded by a single photon with OAM $\ell_{se}$ (where ``$se$'' denotes the seed). The transverse profile of the biphoton down-conversion output state for a SPS-PDC process seeded by a single photon in mode $\tilde{LG}^{\ell_{se}}_{p_{se}}$ in the monochromatic, paraxial, and thin crystal approximations in the low-gain regime is given by \cite{Monken:1998}
\begin{align}
    \label{eqn:StimPDCwf2}
    &|\psi_{\text{SPS-PDC}}\rangle = |0 \rangle |1; \mathbf{q}_{se}, (\ell_{se}, p_{se})\rangle + C \int \! \! \! \int d^2\mathbf{q}_i d^2\mathbf{q}_s \tilde{W}_d (\mathbf{q}_i + \mathbf{q}_s) \hat{a}^\dagger(\mathbf{q}_i) \hat{a}^\dagger(\mathbf{q}_s) \\ \nonumber &\times
    |0; \mathbf{q}_i\rangle |1; \mathbf{q}_{se}, (\ell_{se}, p_{se})\rangle,
\end{align}
where constant $C = \left( \hslash \sqrt{\omega_p \omega_s \omega_i} \chi^{(2)}_{\text{eff}} / \sqrt{\varepsilon_0}\right)$ corresponds to the coefficients in the nonlinear process \cite{Permaul:2025},  $\tilde{W}_d(\cdot)$ corresponds to the Fourier transform of the spatial distribution of the pump at the input face of the crystal and is given by $\tilde{W}_d(\mathbf{q}_d) = (w_d/\sqrt{2\pi}) \exp{\left\{ -w_d^2|\mathbf{q}_i + \mathbf{q}_s|^2/4\right\}}$ where $w_d$ is the beam waist of the pump, $\mathbf{q}$ are the transverse wave vectors of the fields corresponding to the pump ($d$), idler ($i$), signal ($s$), and seed ($se$) respectively, $| 1; \mathbf{q}, (\ell, p)\rangle$ corresponds to a Fock state in mode $\tilde{LG}^\ell_p$ with transverse wavevector $\mathbf{q}$, and $\hat{a}^{(\dagger)}$ denotes an annihilation (creation) operator. The distribution $[N^{\ell_i}_{p_i} ]_{\text{SPS-PDC}}$ corresponds to the mean photon number of idler photons (per temporal detection mode) in mode $\tilde{LG}^{\ell_i}_{p_i}$ and is given by the expectation value $[N^{\ell_i}_{p_i} ]_{\text{SPS-PDC}}= \langle \psi_{\text{SPS-PDC}} | \hat{n}(\mathbf{q}_i ,(\ell_i, p_i))| \psi_{\text{SPS-PDC}} \rangle$ where $\hat{n}(\mathbf{q} ,(\ell, p)) = \hat{a}^\dagger(\mathbf{q} ,(\ell, p))\hat{a}(\mathbf{q} ,(\ell, p))$ represents the number density associated with a photon in mode $\tilde{LG}^\ell_p$ at transverse wavevector $\mathbf{q}$. Here ``per temporal detection mode'' means per gate (for continuous-wave pumping with a window detection window $\Delta t$, opened at frequency $f_{\text{gate}}$) or per pulse (for pulsed pumping at laser repetition frequency $f_{\text{rep}}$); measured count rates follow by multiplying the mean photon number by the gate or laser repetition frequency.  
The first three terms in the binomial expansion of $[N^{\ell_i}_{p_i}] _{\text{SPS-PDC}}$ vanish due to the orthogonality of the Fock states: $\langle n; \mathbf{q}_1| m; \mathbf{q}_2\rangle = \delta(n - m)\delta^2(\mathbf{q}_1 - \mathbf{q}_2)$. Thus, the remaining term in the mean photon number is given by 
\begin{align}
    \label{eqn:C_stimpdc1}
    &[N^{\ell_i}_{p_i}] _{\text{SPS-PDC}} = |C|^2 \int \!\! \cdots \!\!\int d^2\mathbf{q}_{i} d^2\mathbf{q'}_{i} d^2\mathbf{q''}_{i} d^2\mathbf{q'''}_{i} d^2\mathbf{q}_{s} d^2\mathbf{q'}_{s} d^2\mathbf{q}_{se} d^2\mathbf{q'}_{se} \tilde{W}_{p} (\mathbf{q}_{i} + \mathbf{q}_{s}) \\
    &\times  \tilde{W}^*_{p} (\mathbf{q'}_{i} + \mathbf{q'}_{s}) \left [\tilde{LG}_{p_{i}}^{\ell_{i}}  (\mathbf{q''}_{i}) \right]^* \tilde{LG}_{p_{i}}^{\ell_{i}}  (\mathbf{q'''}_{i}) \left [\tilde{LG}_{p_{se}}^{\ell_{se}}  (\mathbf{q}_{se}) \right]^* \tilde{LG}_{p_{se}}^{\ell_{se}}  (\mathbf{q'}_{se}) \nonumber \\ \nonumber &\times \langle 1;\mathbf{q'}_{i} | \hat{a}^\dagger(\mathbf{q''}_i) \hat{a}(\mathbf{q'''}_i) | 1; \mathbf{q}_{i} \rangle  \langle 1;\mathbf{q'}_{se} | \hat{a}(\mathbf{q'}_{s}) \hat{a}^\dagger(\mathbf{q}_{s}) |1; \mathbf{q}_{se}\rangle.
\end{align}
Due to the ladder-operator action on the Fock states and the orthonormality of the Fock states, the inner product $\langle 1;\mathbf{q'}_{i} | \hat{a}^\dagger(\mathbf{q''}_i)$ $\hat{a}(\mathbf{q'''}_i) | 1; \mathbf{q}_{i} \rangle = \delta^2(\mathbf{q''}_i - \mathbf{q'}_i) \delta^2(\mathbf{q'''}_i - \mathbf{q}_i)$. We use the canonical commutation relation in the transverse-momentum continuum $[\hat a(\mathbf{q'}_s),\,\hat a^\dagger(\mathbf q_s)] \;=\; \delta^{(2)}(\mathbf{q'}_s -\mathbf q_s)$ from which it follows that
\begin{align}
    \label{eqn:commutation1}
    & \hat{a}(\mathbf{q'}_{s})\hat{a}^\dagger(\mathbf{q}_{s}) = \delta^2(\mathbf{q'}_{s} - \mathbf{q}_{s}) + \hat{a}^\dagger(\mathbf{q}_{s}) \hat{a}(\mathbf{q'}_{s}).
\end{align}
This relation can be used to evaluate the inner product as
\begin{align}
    &\langle 1;\mathbf{q'}_{se} | \hat{a}(\mathbf{q'}_{s}) \hat{a}^\dagger(\mathbf{q}_{s}) |1; \mathbf{q}_{se}\rangle = \delta^2(\mathbf{q'}_{se} - \mathbf{q}_{se}) \delta^2(\mathbf{q'}_{s} - \mathbf{q}_{s}) + \delta^2(\mathbf{q'}_{se} - \mathbf{q}_{s}) \delta^2(\mathbf{q'}_{s} - \mathbf{q}_{se}).
\end{align}
Due to the orthonormality of the LG modes in $k$-space, $\int d^2 \mathbf{q} \tilde{LG}^{\ell}_{p}(\mathbf{q})\left [\tilde{LG}_{p'_{}}^{\ell'_{}}  (\mathbf{q}_{}) \right]^* = \delta_{\ell,\ell'} \delta_{p,p'}$ the mean photon number evaluates to
\begin{align}
    \label{eqn:P_Stimpdc3}
    &[N^{\ell_i}_{p_i}] _{\text{SPS-PDC}} = |C|^2 \int \!\!\!\int \!\!\!\int d^2\mathbf{q}_i d^2\mathbf{q'}_i d^2\mathbf{q}_s \tilde{W}_d(\mathbf{q}_i + \mathbf{q}_s) \tilde{W}^*_d(\mathbf{q'}_i + \mathbf{q}_s) \tilde{LG}^{\ell_i}_{p_i}(\mathbf{q}_i) \\
    &\times \left[ \tilde{LG}^{\ell_i}_{p_i}(\mathbf{q'}_i) \right]^* + \nonumber \left|C \int \!\!\!\int d^2\mathbf{q}_i d^2\mathbf{q}_s \tilde{W}_d(\mathbf{q}_i + \mathbf{q}_s) \tilde{LG}^{\ell_i}_{p_i}(\mathbf{q}_i) \tilde{LG}^{\ell_{se}}_{p_{se}}(\mathbf{q}_s) \right|^2 .
\end{align}
We note that there exists a weighting factor on the second term in Eq.~(\ref{eqn:P_Stimpdc3}) relative to the first term that depends on the average photon number in the seeding mode. However, since the seeding mode is composed of a single photon, this weighting factor evaluates to 1 which agrees with previous work that shows the weighting factor of the stimulated contribution scales linearly with the average number of photons in the seeding mode \cite{Liscidini:2013}. The overlap integrals in Eq.~(\ref{eqn:P_Stimpdc3}), where $\mathcal{N}^{\ell_i}_{p_i}$ and $|\mathcal{M}^{\ell_i, \ell_{se}}_{p_i, p_{se}}|^2$ correspond to the first and second integrals respectively, have been solved analytically in the Supplementary Information sections (S1) and (S2). This allows for the mean photon number to be expressed as
\begin{equation}
    \label{eqn:P_Stimpdc4}
    [N^{\ell_i}_{p_i}] _{\text{SPS-PDC}}  =  \mathcal{N}^{\ell_i}_{p_i} + \delta_{\ell_i,-\ell_{se}} \left|\mathcal{M}^{\ell_i, \ell_{se}}_{p_i, p_{se}}\right|^2.
\end{equation}
The terms $\mathcal{N}^{\ell_i}_{p_i}$ and $\mathcal{M}^{\ell_i, \ell_{se}}_{p_i, p_{se}}$ are given by the analytical expressions Eq.~(S20) and (S22) and have been plotted in Figs. (\!\!\hyperref[fig:spiralbandwidth]{~\ref{fig:spiralbandwidth}a}) and (\!\!\hyperref[fig:spiralbandwidth]{~\ref{fig:spiralbandwidth}b}) respectively (where the beam waist of the seed and detection beam waists are equal to the beam waist of the pump $w_d = w_{se}= w_i = w'_i = w_s$). The term $\mathcal{N}^{\ell_i}_{p_i}$ corresponds to the mean number of idler photons in mode $\tilde{LG}^{\ell_i}_{p_i}$ in the SPDC process, while the term $\mathcal{M}^{\ell_i, \ell_{se}}_{p_i, p_{se}}$ corresponds to the mean number of signal-idler photon pairs in modes $\tilde{LG}^{\ell_{se}}_{p_{se}}$ and $\tilde{LG}^{\ell_i}_{p_i}$. The origin of both of these terms are explored in the Supplementary Information sections (S3) and (S4). 
We note that the numerical evaluation of $\mathcal{M}^{\ell}_{p_i, p_{se}}$, where $\ell=\ell_i=-\ell_{se}$, shows excellent agreement with the numerical evaluation of the analytical solution of the spiral spectrum of the unseeded SPDC \cite{Miatto:2011} when both expressions are normalized, as plotted in Fig.~(\!\hyperref[fig:spiralbandwidth]{~\ref{fig:spiralbandwidth}c}). Reproducing this distribution in our stimulated scheme indicates that the process implements the expected azimuthal phase conjugation of the seed photon through the underlying OAM conservation in nonlinear optical processes. In the case of perfect phase conjugation in the SPS-PDC process, we require $\mathcal{N}^{\ell_i}_{p_i} = 0$ since it is not constrained by the Kronecker delta $\delta_{\ell_i,-\ell_{se}}$. As a result, a seeding photon carrying OAM $\ell_{se}$ always generates an idler with OAM $\ell_i = -\ell_{se}$. The Kronecker delta $\delta_{\ell_i, -\ell_{se}}$ is written explicitly in Eq.~(\ref{eqn:P_Stimpdc4}) to emphasize that when the stimulated process occurs the idler azimuthal index equals that of the seed. The first term in Eq.~(\ref{eqn:P_Stimpdc4}), $\mathcal{N}^{\ell_i}_{p_i}$, represents the spontaneous contribution to the mean photon number distribution. This term arises because the transverse wave vectors of the signal and seed photons do not interact, meaning that no stimulated process occurs, and the contribution is purely spontaneous. The second term in the sum refers to the stimulated contribution to the coincidence amplitude since in this term, the transverse wave vector of the idler is equal to the conjugate of the seed, as constrained by the Kronecker deltas. It is evident that the key feature indicating phase conjugation in the LG-resolved mean idler photon number distribution is the anti-diagonal along $\ell_i = -\ell_{se}$.

\subsection{Measuring without heralding}
\begin{figure}
    \centering
    \includegraphics[width=0.7\linewidth]{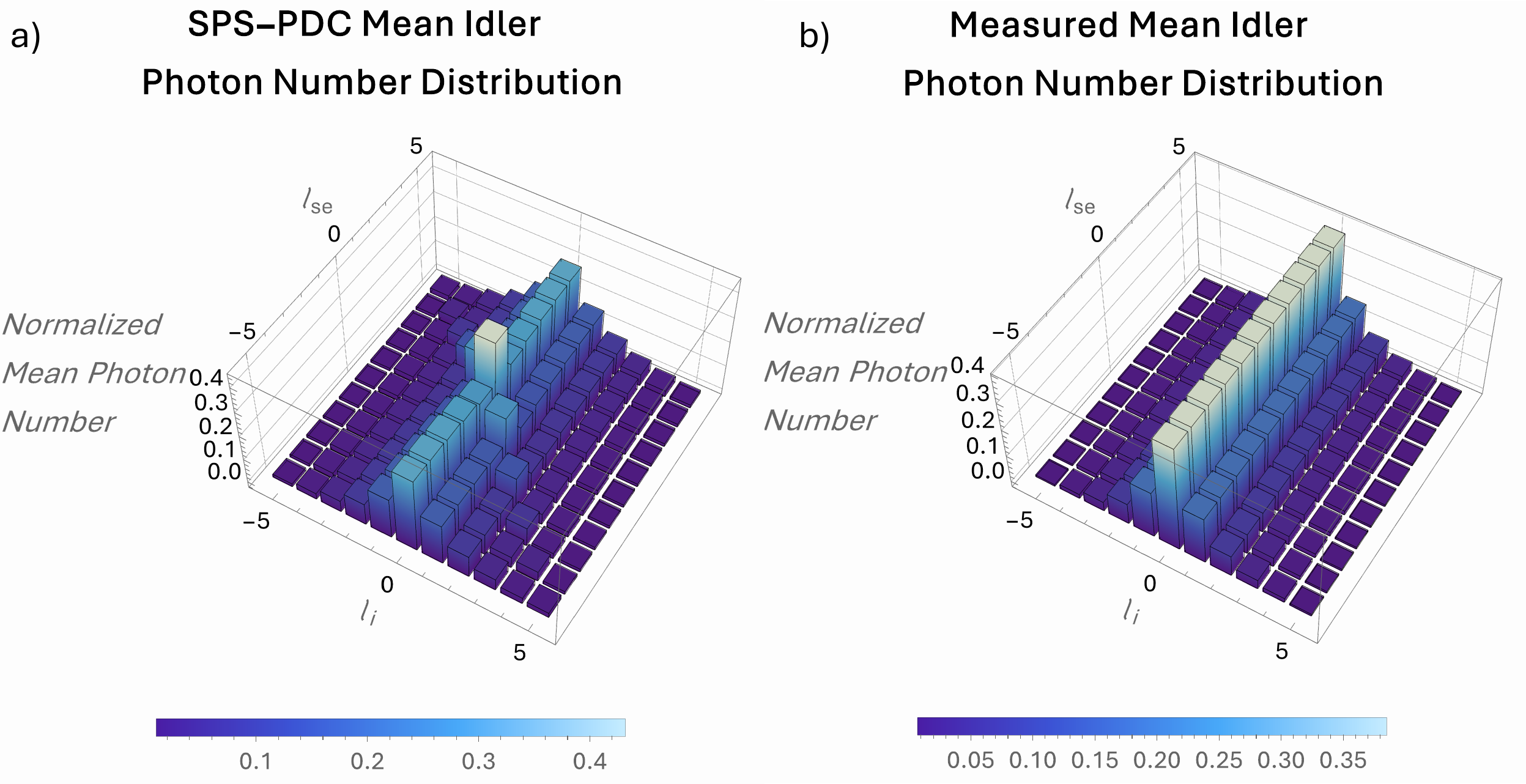}
    \caption{A direct comparison of the simulated normalized mean photon numbers for idler photons in mode $\tilde{LG}^{\ell_i}_{0}$ for a) SPS-PDC, $[N^{\ell_i}_{0}] _{\text{SPS-PDC}}$; b) the measured SPS-PDC scenario with no heralding, $[N^{\ell_i}_{0}] _{\text{measured}}$. As we can observe, the slightly enhanced anti-diagonal OAM correlation pattern that is discernable in a), indicating phase conjugation, becomes heavily suppressed in b).}
    \label{fig:measurementcomparison}
\end{figure}

We now consider what the statistics would look like when experimentally measuring the mean idler photon number in mode $\tilde{LG}^{\ell_i}_{p_i}$ as a function of azimuthal index $\ell_i$. One must consider that SPS-PDC only takes place when both a pump photon and a seed photon are present at the nonlinear crystal within the interaction time of the nonlinear crystal, $\Delta t$. If there is only a pump photon and no seed photon at the nonlinear crystal over $\Delta t$, SPDC can occur. As a result, the measured mean idler photon number will have two contributing terms: a SPDC contribution and a SPS-PDC contribution, each weighted by their respective per-gate occurrence probabilities $\sigma_{\text{SPDC}}$ and $\sigma_{\text{SPS-PDC}}$ which both range from 0 to 1 depending on their relative likelihoods to occur. The measured mean photon number for idler photons in mode $\tilde{LG}^{\ell_i}_{p_i}$ can be expressed as
\begin{align}
    \label{eqn:P_meas1}
    &[N^{\ell_i}_{p_i}] _{\text{measured}} = \sigma_{\text{SPDC}} [N^{\ell_i}_{p_i}] _{\text{SPDC}} + \sigma_{\text{SPS-PDC}} [N^{\ell_i}_{p_i}] _{\text{SPS-PDC}},
\end{align}
where $[N^{\ell_i}_{p_i}] _{\text{SPDC}}$ is the mean photon number for idler photons in LG mode $\tilde{LG}^{\ell_i}_{p_i}$ in SPDC summed over all idler radial indices $p_i$. To see how Eq. (\ref{eqn:P_meas1}) arises, consider the two extremes. The first extreme: no seeding photons are sent and a pump photon is always present at the nonlinear crystal ($R_p \Delta t \gg 1$ and $R_{se} = 0$), so $\sigma_{\text{SPS-PDC}} = 0$ and $\sigma_{\text{SPDC}} = 1$. It is trivial to see how the measured mean idler photon number will correspond to the mean idler photon number in the SPDC process as only SPDC can occur. For the other extreme, assume that there is always a pair of seed and pump photons present at the nonlinear crystal ($R_p \Delta t \gg 1$ and $R_{se} \Delta t \gg 1$). Interactions seeded by vacuum (where a pump photon is present while no seed photon is present) will not occur, so $\sigma_{\text{SPS-PDC}} = 1$ and $\sigma_{\text{SPDC}} = 0$. Similarly, it is clear that the measured mean idler photon number will correspond to the mean idler photon number in the SPS-PDC process, as it is the only process that can occur. So in the ideal scenario, to experimentally measure the distribution $[N^{\ell_i}_{p_i}] _{\text{SPS-PDC}}$ one would require that $\sigma_{\text{SPS-PDC}} = 1$ and $\sigma_{\text{SPDC}} = 0$. However, without heralding the presence of the seed this condition is difficult to satisfy; this is shown in the following calculation. We assume Poissonian statistics to model the arrival probability of both pump and seed photons. While quantum single-photon sources exhibit sub-Poissonian photon-number statistics, the Poissonian approximation captures the probabilistic nature of photon arrival within the interaction window $\Delta t$, which is sufficient for estimating interaction likelihoods \cite{Tey:2025}. The weighting factor $\sigma_{\text{SPDC}}$ then depends on the product of the Poissonian statistics associated with at least one pump photon and no seed photons (vacuum) present over the interaction time $\Delta t$ at the nonlinear crystal
\begin{equation}
    \label{eqn:P_SPDC}
    \sigma_{\text{SPDC}} = (1 - e^{-R_p \Delta t})e^{-R_{se} \Delta t},
\end{equation}
where $R_p$ ($R_{se}$) corresponds to the rate of pump (seed) photons per second. The probability of a stimulated interaction occurring is the joint probability that both a pump and a seed photon are present at the nonlinear crystal over time $\Delta t$
\begin{equation}
    \label{eqn:P_StimPDC}
    \sigma_{\text{SPS-PDC}} = (1 - e^{-R_p \Delta t})(1 - e^{-R_{se} \Delta t}).
\end{equation}
Since most single-photon sources emit single-photons at low rates ($\leq 1.2\times10^8$ Hz) \cite{Uppu:2020, Somaschi:2016, Madsen:2014} and that interaction times in PDC experiments are typically on the order of picoseconds \cite{Slattery:2019, Pomarico:12} (or much shorter in the absence of cavities), we can assume $R_{se}\Delta t\ll 1$ and expand both terms to the first order: $\sigma_{\text{SPDC}} \approx (1 - \exp\{-R_p \Delta t\})(1 - R_{se}\Delta t)$ and $\sigma_{\text{SPS-PDC}} \approx (1 - \exp\{-R_p \Delta t\})R_{se}\Delta t$. Thus, the ratio between the two quantities is
\begin{equation}
    \label{eqn:N_ratio}
    \frac{\sigma_{\text{SPS-PDC}}}{\sigma_{\text{SPDC}}} \approx \frac{R_{se}\Delta t}{1 - R_{se}\Delta t} \ll 1.
\end{equation}
Therefore, the vast majority of the SPS-PDC interaction will be overshadowed by the immense contribution of SPDC to the measured mean idler photon number distribution. This is shown further in Fig.~(\ref{fig:measurementcomparison}) that depicts the mean photon number for idler photons in mode $\tilde{LG}^{\ell_i}_{0}$ (radial indices $p_i = p_{se} = 0$) over idler azimuthal index $\ell_i$ for detection beam waists equal to the beam waist of the pump ($w_d = w_i = w'_i = w_s$) in the SPS-PDC case, Eq.~(\ref{eqn:P_Stimpdc4}), and the measured case, Eq.~(\ref{eqn:P_meas1}) for $\ell_{se}\in[-5,5]$. Both the SPS-PDC distribution Fig.~(\!\hyperref[fig:measurementcomparison]{~\ref{fig:measurementcomparison}a}) and measured distribution  Fig.~(\!\hyperref[fig:measurementcomparison]{~\ref{fig:measurementcomparison}b}) are normalized over all possible idler azimuthal indices $\ell_i$ for each value of $\ell_{se}$, and values of $R_p = 2 \times10^{16}$ photons per second (corresponding to 10 mW of 405 nm), $R_{se}=1.2\times10^8$ photons per second, and $\Delta t = 10$ ps are used in the simulated distributions. In the SPS-PDC case, the OAM of the seed slightly increases the mean photon number when $\ell_i = -\ell_{se}$, which is the signature of phase conjugation. This effect is difficult to observe in the measured case since the number of stimulated events is far outweighed by the number of spontaneous events.
Even with temporal gating, measurements of down-converted modes contingent on the excitation of a single-photon source remain polluted by excess SPDC contributions, as currently available single-photon sources are not 100\% efficient \cite{Uppu:2020, Somaschi:2016, Madsen:2014}. In other words, since not all source excitations produce a photon, many unseeded events (SPDC) appear indistinguishable from genuinely seeded interactions (SPS-PDC). In the following section, we demonstrate that by cascading PDC processes one can overcome the excess SPDC contributions to the measured mean idler photon number distribution through the use of heralding the presence of the seeding photon, allowing the phase conjugation effect in SPS-PDC to be better observed.

\section{Cascaded parametric down-conversion}
\subsection{Joint mean photon number distribution for idler photons in Laguerre-Gaussian modes}
The CPDC configuration, shown in Fig.~(\!\hyperref[fig:StimPDC]{~\ref{fig:StimPDC}b}), heralds the presence of a seeding photon and the occurrence of a subsequent SPS-PDC event using temporal correlations between down-converted photon pairs from an initial SPDC process. This scheme filters excess background SPDC events, isolating only those interactions truly seeded by a single photon. To model this configuration, we develop a theoretical framework describing the expected spatial and temporal correlations between the idler photons when one of the spontaneously down-converted signal photons is used to seed a subsequent SPS-PDC process. In this section, we compute the joint mean photon number distribution for the idler photons, one from each nonlinear interaction, in modes $\tilde{LG}^{\ell_{i1}}_{p_{i1}}$ and $\tilde{LG}^{\ell_{i2}}_{p_{i2}}$, respectively, denoted by $[N^{\ell_{i1}, \ell_{i2}}_{p_{i1}, p_{i2}}]_{\text{CPDC}}$ where the subscripts 1 and 2 denote the first and second nonlinear processes. The resulting distribution contains two distinct terms: one arising from spontaneous emission and one from stimulated emission seeded by a single photon. The OAM representation of the biphoton SPDC output state in the monochromatic, paraxial, and thin crystal approximations is given by \cite{Torres:2003}
\begin{align}
    \label{eqn:SPDCstate}
    &|\psi_{\text{SPDC}}\rangle = \sum_{\ell_i, \ell_s = -\infty}^\infty \sum_{p_i, p_s = 0}^\infty  \mathcal{M}^{\ell_i, \ell_s}_{p_i,p_s} \int d^2 \mathbf{q}_s \left[ \tilde{LG}^{\ell_s}_{p_s}(\mathbf{q}_s)\right]^* |1;\mathbf{q}_{i},(\ell_i, p_i)\rangle |1;\mathbf{q}_s \rangle,
\end{align}
where $\mathcal{M}^{\ell_i, \ell_s}_{p_i, p_s}$ are the coincidence amplitudes associated with observing a pair of idler-signal photons in modes $\tilde{LG}^{\ell_i}_{p_i}$ and $\tilde{LG}^{\ell_s}_{p_s}$ in the output of SPDC, otherwise known as the spiral spectrum \cite{Miatto:2011}; a derivation of this relation is developed in Supplementary Information section (S4). The azimuthal portion of this integral introduces the Kronecker delta $\delta_{\ell_i,-\ell_s}$; thus, without loss of generality, we homogenize the azimuthal indices such that $\ell_i = -\ell_s = \ell$. The resulting quad-photon CPDC output state is then found by substituting Eq.~(\ref{eqn:SPDCstate}) as the seeding state in the typical StimPDC output state in the monochromatic, paraxial, and thin crystal approximations and the low-gain regime \cite{Riberiro:1999}
\begin{align}
    \label{eqn:CPDC1}
    &|\psi_{\text{CPDC}} \rangle =  \sum_{\ell = -\infty}^\infty \sum_{p_i, p_s = 0}^\infty  \mathcal{M}^{\ell}_{p_i,p_s}\left[ \int d^2\mathbf{q}_{s1} \left[ \tilde{LG}^{-\ell}_{p_s}(\mathbf{q}_{s1})\right]^* |1;\mathbf{q}_{i1},(\ell, p_i)\rangle |1; \mathbf{q}_{s1}\rangle \right.
    \\& \nonumber + C_2 \int \! \! \! \int \! \! \! \int d^2\mathbf{q}_{s1} d^2\mathbf{q}_{i2} d^2\mathbf{q}_{s2} \tilde{W}_{d2}(\mathbf{q}_{i2} + \mathbf{q}_{s2}) \left[ \tilde{LG}^{-\ell}_{p_s}(\mathbf{q}_{s1})\right]^*\hat{a}^\dagger(\mathbf{q}_{s2})\\ &\times \left. |1; \mathbf{q}_{i1}, (\ell, p_i) \rangle |1; \mathbf{q}_{s1} \rangle |1; \mathbf{q}_{i2} \rangle \right].\nonumber 
\end{align}
The joint mean photon number distribution associated with idler photons in modes $\tilde{LG}^{\ell_{i1}}_{p_{i1}}$ and $\tilde{LG}^{\ell_{i2}}_{p_{i2}}$ (per temporal detection mode) are given by $[N^{\ell_{i1}, \ell_{i2}}_{p_{i1}, p_{i2}}]_{\text{CPDC}} = \langle \psi_{\text{CPDC}}| \hat{n}(\mathbf{q}_{i1} ,(\ell_{i1}, p_{i1}))$ $ \hat{n}(\mathbf{q}_{i2} ,(\ell_{i2}, p_{i2}))| \psi_{\text{CPDC}} \rangle$. 
Expanding the expectation value results in a binomial expansion; the first three terms of this expansion vanish due to the orthogonality of the Fock states. We again impose the relation in Eq.~(\ref{eqn:commutation1}) and the ladder-operator action on Fock states, together with the orthonormality of the Fock basis, to evaluate the matrix elements. This introduces the Kronecker deltas $\delta_{\ell',\ell_{i1}} \delta_{\ell_{i1},\ell} \delta_{p'_i,p_{i1}} \delta_{p_{i1},p_i}$ and Dirac delta functions $\delta^2(\mathbf{q'}_{i1} - \mathbf{q''}_{i1}) \delta^2(\mathbf{q'''}_{i1} - \mathbf{q}_{i1})$ for the first inner product, which collapses the sums over $\ell$, $\ell'$ and $p_i$, $p'_i$ at $\ell_{i1}$ and $p_{i1}$, and the integrals over $\mathbf{q''}_{i1}$ and $\mathbf{q'''}_{i1}$. Similarly, the second inner product collapses the integrals over $\mathbf{q''}_{i2}$ and $\mathbf{q'''}_{i2}$. The orthogonality of the LG modes forces $p_s = p'_s$ in the first term. This simplifies $[N^{\ell_{i1}, \ell_{i2}}_{p_{i1}, p_{i2}}]_{\text{CPDC}}$ as
\begin{align}
    \label{eqn:P_CPDC2}
    &[N^{\ell_{i1}, \ell_{i2}}_{p_{i1}, p_{i2}}]_{\text{CPDC}} = \sum_{p_s = 0}^\infty  |\mathcal{M}^{\ell_{i1}}_{p_{i1},p_s}|^2 |C_2|^2 \int \!\!\! \int \!\!\! \int d^2 \mathbf{q}_{i2} d^2 \mathbf{q'}_{i2} d^2 \mathbf{q}_{s2} \tilde{W}_{d2}(\mathbf{q}_{i2} + \mathbf{q}_{s2}) \\
    &\times \nonumber \tilde{W}^*_{d2}(\mathbf{q'}_{i2} + \mathbf{q}_{s2}) \tilde{LG}^{\ell_{i2}}_{p_{i2}}(\mathbf{q}_{i2}) \left[\tilde{LG}^{\ell_{i2}}_{p_{i2}}(\mathbf{q'}_{i2}) \right]^* + \sum_{p_s, p'_s = 0}^\infty C_2^* \int \!\!\! \int d^2 \mathbf{q}_{s1} \\
    &\times \nonumber d^2 \mathbf{q'}_{i2} \tilde{W}^*_{d2}(\mathbf{q}_{s1} + \mathbf{q'}_{i2}) \left[\tilde{LG}^{-\ell_{i1}}_{p_s}(\mathbf{q}_{s1}) \right]^* \left[\tilde{LG}^{\ell_{i2}}_{p_{i2}}(\mathbf{q'}_{i2}) \right]^*  \\ \nonumber &\times C_2\int \!\!\! \int d^2 \mathbf{q}_{i2} d^2 \mathbf{q}_{s2} \tilde{W}_{d2}(\mathbf{q}_{i2} + \mathbf{q}_{s2}) \tilde{LG}^{\ell_{i2}}_{p_{i2}}(\mathbf{q}_{i2}) \tilde{LG}^{-\ell_{i1}}_{p'_s}(\mathbf{q}_{s2}).
\end{align}
We then note that the remaining triple integral in the first term of Eq.~(\ref{eqn:P_CPDC2}) corresponds to the spontaneous overlap integral found in the first term of Eq.~(\ref{eqn:P_Stimpdc3}) which is solved analytically in Eq.~(S20). The second term of Eq.~(\ref{eqn:P_CPDC2}) contains two double integrals that have the same form as the double integrals found in the second term of Eq.~(\ref{eqn:P_Stimpdc3}), with both having the solution found in Eq.~(S22). Each of these double integrals introduces the Kronecker delta $\delta_{\ell_{i1}, \ell_{i2}}$, which is a signature of the stimulating process. The resulting joint mean idler photon number is given by
\begin{align}
    \label{eqn:P_CPDC3}
     &[N^{\ell_{i1}, \ell_{i2}}_{p_{i1}, p_{i2}}]_{\text{CPDC}} = \sum_{p_s = 0}^\infty  \left| \mathcal{M}^{\ell_{i1}}_{p_{i1},p_s} \right|^2 \mathcal{N}^{\ell_{i2}}_{p_{i2}} + \sum_{p_s, p'_s = 0}^\infty  \mathcal{M}^{\ell_{i1}}_{p_{i1},p_s} \left[\mathcal{M}^{\ell_{i1}}_{p_{i1},p'_s} \right]^* \\ \nonumber &\times \left[\mathcal{M}^{-\ell_{i1}, \ell_{i2}}_{p_{s}, p_{i2}}\right]^* \mathcal{M}^{-\ell_{i1}, \ell_{i2}}_{p'_{s}, p_{i2}}.
\end{align}
Notably, when $p_s = p'_s$, Eq.~(\ref{eqn:P_CPDC3}) reduces to the form of Eq.~(\ref{eqn:P_Stimpdc4}), multiplied by the factor $\sum_{p_s = 0}^\infty|\mathcal{M}^{\ell_i}_{p_{i1},p_s}|^2$. This shows that the CPDC configuration does not generate a new OAM distribution beyond the SPS-PDC distribution; the SPS-PDC distribution is reweighed by $\sum_{p_s = 0}^\infty|\mathcal{M}^{\ell_i}_{p_{i1},p_s}|^2$. Similarly to the case of SPS-PDC with an ideal single-photon source, this distribution consist of a spontaneous background contribution and a stimulated contribution that appears when the two idler indices are equal ($\ell_{i1} = \ell_{i2}$). Therefore, mean photon numbers along the diagonal will indicate the presence of the phase conjugation effect at the single-photon level. 

\subsection{The advantage of heralding}
\begin{figure}
    \centering
    \includegraphics[width=0.9\linewidth]{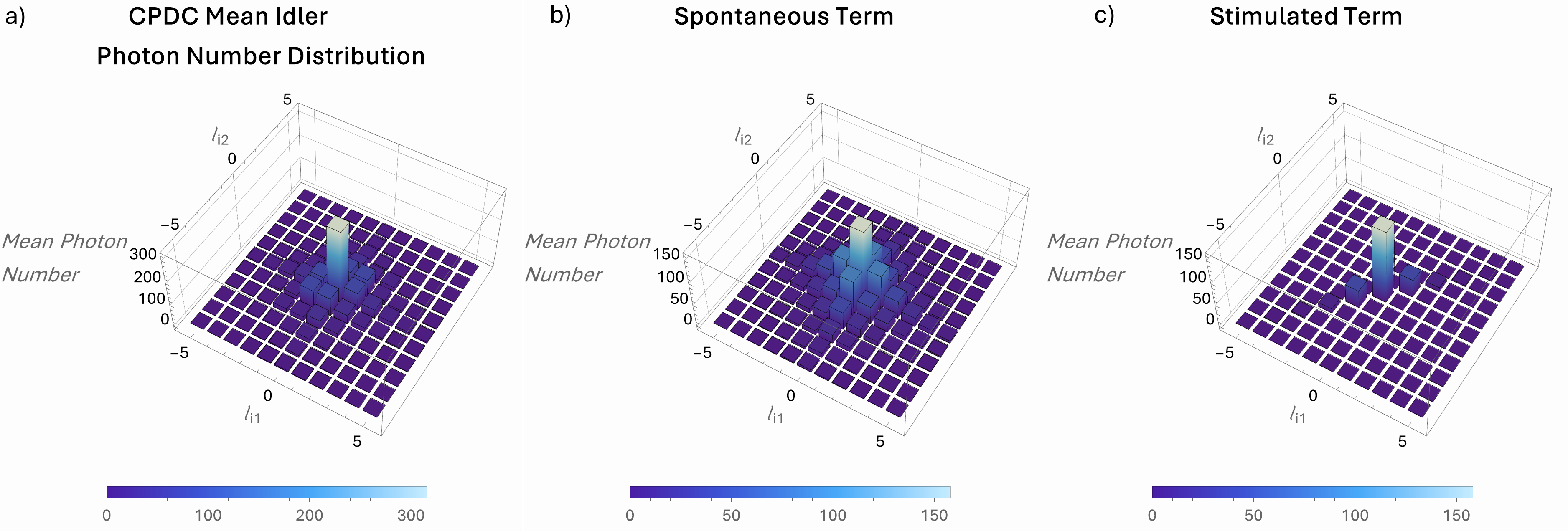}
    \caption{ The joint simulated mean photon numbers associated with generating idler photon pairs with indices $\ell_{i1}, \ell_{i2}$ for the a) CPDC configuration, showing a strong diagonal along $\ell_{i1}=\ell_{i2}$ corresponding to phase conjugation; b) spontaneous contribution, which is broad due to the lack of correlation between the two idler modes c) stimulated contribution, present only along the diagonal $\ell_{i1}=\ell_{i2}$ as the second idler is produced in the conjugate OAM mode of the seeding photon ($-\ell_{s1}=\ell_{i1}$). The mean photon number distribution in a) is obtained by summing the spontaneous and stimulated terms.}
    \label{fig:CPDC_plot}
\end{figure}

This cascaded configuration offers a key advantage over the single-photon-seeded StimPDC: the experimentally measured quantity $[N^{\ell_{i1}, \ell_{i2}}_{p_{i1}, p_{i2}}]_{\text{CPDC}}$ heralded by detections of idler photons in specified LG modes directly corresponds to the theoretical mean photon number distribution $[N^{\ell_{i1}, \ell_{i2}}_{p_{i1}, p_{i2}}]_{\text{CPDC}}$. In this setup, coincidences are recorded only when both idler photons are detected within the correlation window, effectively ensuring that the SPS-PDC emission is measured exclusively when a seeding photon is present. This approach suppresses contributions from purely spontaneous interactions, which otherwise dominate in the measured SPS-PDC configuration. 
The enhancement of the phase conjugation signature in CPDC can be exemplified when plotting the joint mean idler photon number distribution given by Eq. (\ref{eqn:P_CPDC3}). The plot Fig.~(\!\hyperref[fig:CPDC_plot]{~\ref{fig:CPDC_plot}a}) shows the joint simulated mean photon number distribution $[N^{\ell_{i1}, \ell_{i2}}_{0, 0}]_{\text{CPDC}}$ for generating idler photon pairs in the CPDC configuration ($p_{i1} = p_{i2} = 0$). The distribution exhibits an enhancement along the diagonal $\ell_{i1}=\ell_{i2}$, indicating the phase conjugation effect. To illustrate the origin of this feature, we separate the total distribution into its spontaneous and stimulated contributions in Figs. (\!\hyperref[fig:CPDC_plot]{~\ref{fig:CPDC_plot}b}) and (\!\hyperref[fig:CPDC_plot]{~\ref{fig:CPDC_plot}c}) respectively. The spontaneous term lacks an enhancement along the diagonal, consistent with the absence of transverse-mode interaction between the seed and down-converted photons in the second nonlinear crystal. In contrast, the stimulated term is confined to the diagonal $\ell_{i1}=\ell_{i2}$ due to the Kronecker delta $\delta_{\ell_{i1},\ell_{i2}}$ reflecting the constraint $\ell_{i2}=-\ell_{s1}=\ell_{i1}$ imposed by the interaction between the seed's transverse wavevector and the down-converted modes.

Both Figs. (\!\!\hyperref[fig:CPDC_plot]{~\ref{fig:CPDC_plot}c}) and (\!\!\hyperref[fig:spiralbandwidth]{~\ref{fig:spiralbandwidth}b}) illustrate the same single-photon-level phase conjugation process; however, their spectral widths differ. In Fig.~(\!\hyperref[fig:spiralbandwidth]{~\ref{fig:spiralbandwidth}b}), we assume a uniform seeding probability of one photon for all $\ell$ values. In contrast, for the CPDC case shown in Fig.~(\!\!\hyperref[fig:CPDC_plot]{~\ref{fig:CPDC_plot}c}), the probability of seeding a photon varies and decreases depending on its $\ell_i$ value. Previous studies have shown that increasing the beam waist ratio between the pump and the seed modifies the single-photon probability distribution as a function of $\ell_i$ \cite{Xu:2024}. This adjustment could, in principle, be used to reshape the distribution observed in Fig.~(\!\!\hyperref[fig:CPDC_plot]{~\ref{fig:CPDC_plot}c}); this effect is briefly explored in Supplementary Information section (S5). 

\begin{figure}
    \centering
    \includegraphics[width=0.8\linewidth]{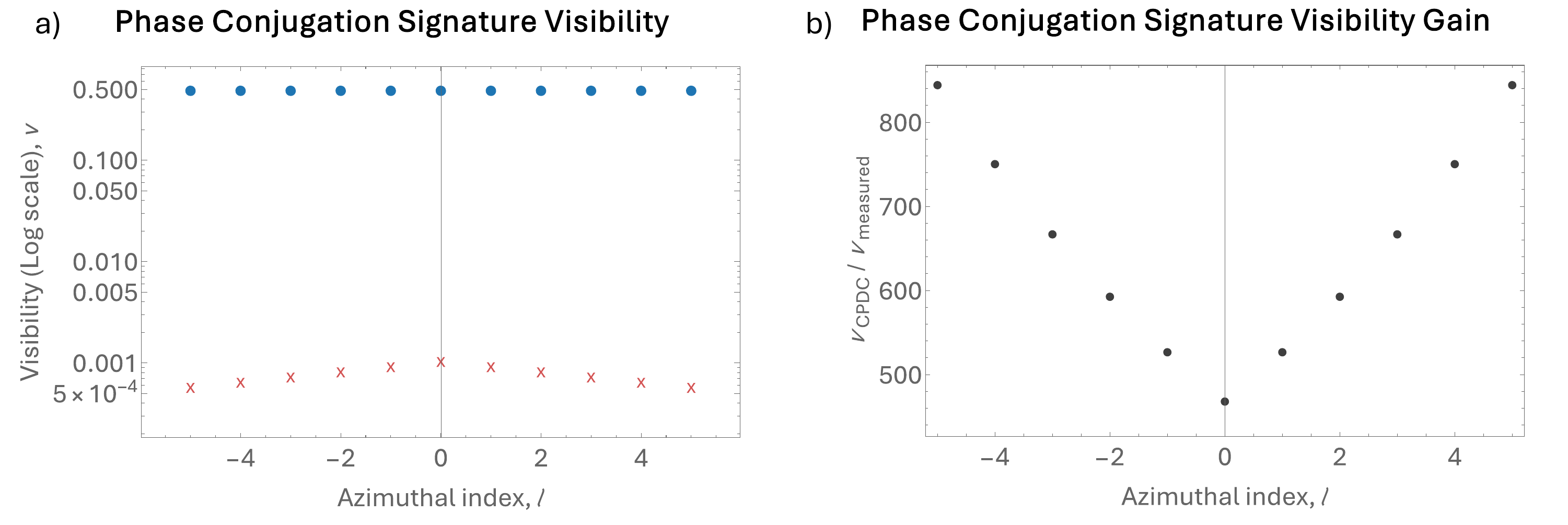}
    \caption{a) A comparison between the visibilities of the phase conjugation signature for the CPDC distribution (blue dots) and measured SPS-PDC distribution (red crosses); b) the visibility gain between the CPDC and measured SPS-PDC configurations. Heralding the presence of the seeding photon allows the visibility of the phase conjugation signature in the CPDC configuration to be heavily enhanced relative to the measured SPS-PDC configuration.}
    \label{fig:visibilities}
\end{figure}

To quantify the advantage of using the CPDC configuration, we characterize the visibility of the phase conjugation signature in both the CPDC and measured SPS-PDC mean photon number distributions. Since phase conjugation only occurs along specific pairs of azimuthal indices in both cases, along the diagonal $\ell_{i1} = \ell_{i2}$ in the CPDC case and along the anti-diagonal $\ell_{i} = -\ell_{se}$ in the SPS-PDC case, the visibility will only be considered for these indices ($\ell = \ell_{i1} = \ell_{i2} = \ell_i = -\ell_{se}$). Additionally, we will examine the case where $p_{i} = p_{se} = p_{i1} = p_{i2} = 0$. The visibility of the phase conjugation signature can be quantified for each index $\ell$ as the fraction of the mean photon number that the stimulated term contributes
\begin{equation}
    \label{eqn:CPDC_vis}
    [\nu^\ell_{0, 0}]_{\text{CPDC}} = \frac{\sum_{p_s, p'_s = 0}^\infty  \mathcal{M}^{\ell}_{0,p_s} \left[\mathcal{M}^{\ell}_{0,p'_s} \right]^*   \left[\mathcal{M}^{\ell}_{p_{s}, 0}\right]^* \mathcal{M}^{\ell}_{p'_{s}, 0}}{[N^{\ell}_{0, 0}]_{\text{CPDC}}},
\end{equation}
for the CPDC distribution and 
\begin{equation}
    \label{eqn:SPS-PDC_vis}
    [\nu^\ell_0]_{\text{measured}} = \frac{\sigma_{\text{SPS-PDC}}\left|\mathcal{M}^{\ell}_{0, 0}\right|^2}{ [N^{\ell}_{0}]_{\text{measured}}},
\end{equation}
for the SPS-PDC distribution. The visibilities $[\nu^\ell_{0, 0}]_{\text{CPDC}}$ and $[\nu^\ell_{0}]_{\text{measured}}$ are plotted in Fig.~(\!\hyperref[fig:visibilities]{~\ref{fig:visibilities}a}) where a logarithmic scale is used due to the difference in magnitude between the two quantities. Furthermore, the visibility gain in the CPDC configuration, $[\nu^\ell_{0, 0}]_{\text{CPDC}}/[\nu^\ell_0]_{\text{measured}}$, is plotted in Fig.~(\!\hyperref[fig:visibilities]{~\ref{fig:visibilities}b}). The higher visibilities in the CPDC configuration indicate that heralding the presence of the seeding photon allows for the phase conjugation signature to be observed much easier. This is further supported by the visibility gain of two orders of magnitude for all values of $\ell$ in the CPDC configuration relative to the measured SPS-PDC configuration.

\section{Conclusion}
We have developed a fully analytical model for CPDC that captures both the spontaneous and stimulated contributions to the mean photon number distribution $[N^{\ell_{i1}, \ell_{i2}}_{p_{i1}, p_{i2}}]_{\text{CPDC}}$ associated with generating idler photon pairs in specific LG modes in CPDC. The model predicts that the spontaneous term yields a broad background with no enhanced diagonal structure, whereas the stimulated term produces a distinct enhancement in the distribution along $\ell_{i1}=\ell_{i2}$. This provides a clear theoretical signature for identifying azimuthal phase conjugation at the single-photon level. The advantage of the CPDC configuration over performing SPS-PDC without heralding for the presence of the seed is the ability to filter out unwanted SPDC contributions, thereby increasing the visibility of the phase-conjugation signature by two orders of magnitude. This approach represents a significant step toward experimentally realizing quantum nonlinear optical effects with structured single photons, opening pathways for advancing quantum information processing schemes. 

\begin{backmatter}

\bmsection{Author contributions}
Conceptualization - AAAC. Methodology - TJBL, AAAC. Formal analysis - TJBL, AAAC. Investigation - TJBL, AAAC. Software - TJBL. Validation - CL, MFFG, JU, JSL, RWB. Visualization - TJBL. Writing - original draft - TJBL. Writing - review \& editing - AAAC, CL, MFFG, JU, JSL, RWB. Supervision - AAAC, MFG. Project administration - MFFG. Funding acquisition - JSL, RWB.
\end{backmatter}

\bibliography{_TEXT}

\clearpage
\appendix
\counterwithin{section}{part}
\renewcommand{\thesection}{S\arabic{section}}
\setcounter{section}{0}

\setcounter{figure}{0}
\renewcommand{\thefigure}{S\arabic{figure}}
\setcounter{table}{0}
\renewcommand{\thetable}{S\arabic{table}}
\setcounter{equation}{0}
\renewcommand{\theequation}{S\arabic{equation}}

\section{SUPPLEMENTARY INFORMATION}
\subsection{Solving the spontaneous LG overlap integral}
\label{sec:LGInt1}
In this section, we analytically solve the overlap integral in the spontaneous term in the sum of Eq. (6) in the main text. Specifically,
\begin{align}
    \label{eqn:A1}
    \mathcal{N}^{\ell_i}_{p_i} = |C|^2 \int \!\!\!\int \!\!\!\int d^2\mathbf{q}_i d^2\mathbf{q'}_i d^2\mathbf{q}_s \tilde{W}_d(\mathbf{q}_i + \mathbf{q}_s) \tilde{W}^*_d(\mathbf{q'}_i + \mathbf{q}_s) \tilde{LG}^{\ell_i}_{p_i}(\mathbf{q}_i) \left[ \tilde{LG}^{\ell_i}_{p_i}(\mathbf{q'}_i) \right]^*.
\end{align}
First, the integral over $\mathbf{q}_s$ can be evaluated,
\begin{align}
    &\int d^2 \mathbf{q}_s \tilde{W}_d(\mathbf{q}_i + \mathbf{q}_s) \tilde{W}^*_d(\mathbf{q'}_i + \mathbf{q}_s) = \frac{w^2_d}{2\pi}\int d^2 \mathbf{q}_s e^{-w_d^2|\mathbf{q}_i + \mathbf{q}_s|^2/4} e^{-w_d^2|\mathbf{q'}_i + \mathbf{q}_s|^2/4} \\
    &= \frac{w^2_d}{2\pi}\int d^2 \mathbf{q}_s e^{-w_d^2(2|\mathbf{q}_s + (\mathbf{q}_i + \mathbf{q'}_i)/2|^2 + |\mathbf{q}_i - \mathbf{q'}_i|^2/2)/4} \nonumber \\
    &= \frac{w^2_d}{2\pi} e^{-w_d^2|\mathbf{q}_i - \mathbf{q'}_i|^2/8}\int d^2 \mathbf{q}_s e^{-w_d^2(|\mathbf{q}_s + (\mathbf{q}_i + \mathbf{q'}_i)/2|^2)/2} \nonumber \\
    &= e^{-w_d^2|\mathbf{q}_i - \mathbf{q'}_i|^2/8} \nonumber.
\end{align}
Now Eq. (\ref{eqn:A1}) can be evaluated as a double integral over $\mathbf{q}_i$ and $\mathbf{q'}_i$
\begin{align}
    \label{eqn:A2}
    &\mathcal{N}^{\ell_i}_{p_i} = |C|^2 \int \!\!\!\int d^2\mathbf{q}_i d^2\mathbf{q'}_i e^{-w_d^2|\mathbf{q}_i - \mathbf{q'}_i|^2/8} \tilde{LG}^{\ell_i}_{p_i}(\mathbf{q}_i) \left[ \tilde{LG}^{\ell_i}_{p_i}(\mathbf{q'}_i) \right]^* \\
    &= \int_{q_i = 0}^\infty dq_i \int_{\phi_i = 0}^{2\pi} q_i d\phi_i \int_{q'_i = 0}^\infty dq'_i \int_{\phi'_i = 0}^{2\pi} q'_i d\phi'_i e^{-w_d^2|\mathbf{q}_i - \mathbf{q'}_i|^2/8} \sqrt{\frac{w_i^2 p_i!}{2\pi(p_i + |\ell_i|)!}} \sqrt{\frac{w'_i\,^2 p_i!}{2\pi(p_i + |\ell_i|)!}}\nonumber \\
    &\times \left( \frac{w_i q_i}{\sqrt{2}} \right)^{|\ell_i|} \left( \frac{w'_i q'_i}{\sqrt{2}} \right)^{|\ell_i|} e^{-w_i^2 q_i^2 / 4} e^{-w'_i\,^2 q'_i\,^2 / 4} (-1)^{p_i + p_i} L_{p_i}^{|\ell_i|}\left(\frac{w_i^2 q_i^2}{2}\right) L_{p_i}^{|\ell_i|}\left(\frac{w'_i\,^2 q'_i\,^{2}}{2}\right) \nonumber \\ &\times e^{i\ell_i(\phi_i + \pi/2)} e^{-i\ell_i(\phi'_i + \pi/2)} \nonumber.
\end{align}
The exponential kernel can be written in polar coordinates by making the substitution $|\mathbf{q}_i - \mathbf{q}'_i|^2 = q_i^2 + q'_i\,^2 - 2q_iq'_i \cos(\phi_i - \phi'_i)$. The kernel now depends on the radial coordinates in momentum space, $q_i$ and $q'_i$, and on the difference between azimuthal angles, $\phi_i - \phi'_i$. We can then write the kernel as a superposition of plane waves with phase $\exp\{i\ell(\phi_i - \phi'_i)\}$ \cite{Miatto:2011}
\begin{equation}
    \label{planewavepumpapprox}
    e^{-w_d^2(q_i^2 + q'_i\,^2 - 2q_iq'_i \cos(\phi_i - \phi'_i))/8} = \sum_{\ell' = -\infty}^\infty f_{\ell'}(q_i, q'_i) e^{i\ell'(\phi_i - \phi'_i)}.
\end{equation}
Where the Fourier coefficient $f_{\ell'}(q_i, q'_i)$ has the form
\begin{align}
    &f_{\ell'}(q_i, q'_i) = \frac{1}{2\pi} \int_0^{2\pi} d\phi \, e^{-w_d^2(q_i^2 + q'_i\,^2 - 2q_i q'_i \cos \phi)/8} e^{i\ell'\phi} \\
    &= \frac{1}{2\pi} e^{-w_d^2(q_i^2 + q'_i\,^2)/8} \int_0^{2\pi} d\phi \, e^{w_d^2 q_i q'_i \cos \phi/4} e^{i\ell'\phi}. \nonumber
\end{align}
This is a standard integral representation of the modified Bessel function of the first kind
\begin{align}
    I_\ell(z) = \frac{1}{2\pi} \int_0^{2\pi} e^{z \cos \phi} e^{i \ell \phi} d\phi,
\end{align}
so we obtain
\begin{align}
    f_{\ell'}(q_i, q'_i) = e^{-w_d^2(q_i^2 + q'_i\,^2)/8} I_{\ell'}\left(\frac{w_d^2 q_i q'_i}{4}\right).
\end{align}
The integral then becomes
\begin{align}
    \label{eqn:A3}
    &\mathcal{N}^{\ell_i}_{p_i} = \frac{|C|^2 w_i w'_i p_i!}{2\pi(p_i + |\ell_i|)!} \left( \frac{w_i w'_i}{2} \right)^{|\ell_i|} \int_{q_i = 0}^\infty \int_{q'_i = 0}^\infty dq_i dq'_i (q_i q'_i)^{|\ell_i| + 1} e^{-w_i^2 q_i^2 / 4} e^{-w'_i\,^2 q'_i\,^2 / 4} L_{p_i}^{|\ell_i|}\left(\frac{w_i^2 q_i^2}{2}\right)
    \\&\times L_{p_i}^{|\ell_i|}\left(\frac{w'_i\,^2 q'_i\,^2}{2}\right) \sum_{\ell' = -\infty}^\infty e^{-w_d^2(q_i^2 + q'_i\,^2)/8} I_{\ell'}\left(\frac{w_d^2 q_i q'_i}{4}\right) \int_{\phi_i = 0}^{2\pi} \int_{\phi'_i = 0}^{2\pi} d\phi_i d\phi'_i e^{i\ell'(\phi_i - \phi'_i)} e^{i\ell_i(\phi_i + \pi/2)} \nonumber \\ &\times e^{-i\ell_i(\phi'_i + \pi/2)} \nonumber.
\end{align}
With the angular integral now expressing the form 
\begin{equation}
    \label{angularint}
    \sum_{\ell' = -\infty}^\infty e^{-w_d^2(q_i^2 + q'_i\,^2)/8} I_{\ell'}\left(\frac{w_d^2 q_i q'_i}{4}\right) \int_{\phi_i = 0}^{2\pi} d\phi_i e^{i\phi_i(\ell' + \ell_i)} e^{i\ell_i \pi/2} \int_{\phi'_i = 0}^{2\pi} d\phi'_i e^{-i\phi'_i(\ell' + \ell_i)} e^{-i\ell_i \pi/2}.
\end{equation}
Due to the orthogonality of exponential functions $\int_0^{2\pi} e^{i\phi} d\phi = 2\pi\delta_\phi$, the angular integral evaluates to
\begin{equation}
    \label{angularint2}
    \sum_{\ell' = -\infty}^\infty e^{-w_d^2(q_i^2 + q'_i\,^2)/8} I_{\ell'}\left(\frac{w_d^2 q_i q'_i}{4}\right) 4 \pi^2 \delta_{\ell',-\ell_i} \delta_{\ell', -\ell_i}.
\end{equation}
Under integer indices, $I_\ell(\cdot) = I_{-\ell}(\cdot) = I_{|\ell|}(\cdot)$. The remaining radial integral then simplifies to
\begin{align}
    \label{eqn:A4}
    \mathcal{N}^{\ell_i}_{p_i} &= \frac{2 \pi |C|^2 w_i w'_i p_i!}{(p_i + |\ell_i|)!} \left( \frac{w_i w'_i}{2} \right)^{|\ell_i|} \int_{q_i = 0}^\infty dq_i e^{-(w_d^2 + 2w_i^2)q_i^2/8} q_i^{|\ell_i| + 1} L_{p_i}^{|\ell_i|}\left(\frac{w_i^2 q_i^2}{2}\right) 
    \\& \times \int_{q'_i = 0}^\infty dq'_i e^{-(w_d^2 + 2w'_i\,^2)q'_i\,^2/8} (q_i')^{|\ell_i| + 1} L_{p_i}^{|\ell_i|}\left(\frac{w'_i\,^2 q'_i\,^2}{2}\right) I_{|\ell_i|}\left(\frac{w_d^2 q_i q'_i}{4}\right)\nonumber.
\end{align}
The term $I_{|\ell_i|}\left(w_d^2 q_i q'_i/4\right)$ can be expanded as a series
\begin{align}
    \label{Seriesexpansion}
    &I_{|\ell_i|}\left(\frac{w_d^2 q_i q'_i}{4}\right) = \sum_{r = 0}^\infty\frac{1}{r! \Gamma(r + |\ell_i| + 1)} \left( \frac{w_d^2 q_i q'_i}{8}\right)^{2r + |\ell_i|} \\& \nonumber = \sum_{r = 0}^\infty \frac{1}{r! \Gamma(r + |\ell_i| + 1)} \left( \frac{w_d^2}{8}\right)^{2r + |\ell_i|} \left(q_i\right)^{2r + |\ell_i|} \left( q'_i\right)^{2r + |\ell_i|}.
\end{align}
This allows for the integrals over $q_i$ and $q'_i$ to be completely decoupled
\begin{align}
    \label{eqn:A5}
    &\mathcal{N}^{\ell_i}_{p_i} = \frac{2\pi |C|^2 w_i w'_i p_i!}{(p_i + |\ell_i|)!} \left( \frac{w_i w'_i}{2} \right)^{|\ell_i|} \sum_{r=0}^{\infty} \frac{(w_d^2 / 8)^{2r + |\ell_i|}}{r! \Gamma(r + |\ell_i| + 1)} \int_{q_i = 0}^\infty dq_i e^{-(w_d^2 + 2w_i^2)q_i^2/8} q_i^{2|\ell_i| +  2r + 1} 
    \\& \times  L_{p_i}^{|\ell_i|}\left(\frac{w_i^2 q_i^2}{2}\right)\int_{q'_i = 0}^\infty dq'_i e^{-(w_d^2 + 2w'_i\,^2)q'_i\,^2/8} q'_i\,^{2|\ell_i| + 2r + 1} L_{p_i}^{|\ell_i|}\left(\frac{w'_i\,^2 q'_i\,^2}{2}\right) \nonumber .
\end{align}
The integrals over $q_i$ and $q'_i$ are identical, aside from the change of indices. These integrals can be evaluated individually; we will now evaluate the general integral. 
\begin{equation}
    \label{eqn:B1}
    \mathcal{B} = \int_{0}^\infty dq e^{-(w_d^2 + 2w^2) q^2/8} q^{2|\ell| + 2r + 1} L_{p}^{|\ell|}\left( \frac{w^2 q^2}{2}\right).
\end{equation}
Substituting $x = q^2 w^2 / 2$ such that $dq = dx/(q w^2)$, and $q^{2|\ell| + 2r} = (2/w^2)^{|\ell| + r} x^{|\ell| + r}$
\begin{equation}
    \label{eqn:B2}
    \mathcal{B} = \int_{0}^\infty dx \frac{1}{w^2} \left(\frac{2}{w^2}\right)^{|\ell| + r} e^{-x(w_d^2 + 2w^2)/4w^2} x^{|\ell| + r} L_{p}^{|\ell|}(x).
\end{equation}
Adapting Gradshteyn $\&$ Ryzhik 7.414.7 \cite{Gradshteyn:2015}
\begin{equation}
    \label{Gradshteyn}
    \int_0^\infty e^{-st} t^\beta L_n^\alpha(t) dt = \frac{\Gamma(\beta + 1)\Gamma(\alpha + n + 1)}{n! \Gamma(\alpha + 1)}s^{-\beta - 1} {}_{2}F_{1}\left[-n, \beta + 1; \alpha + 1; \frac{1}{s}\right],
\end{equation}
which converges for $\text{Re}(\beta) > -1$ and $\text{Re}(s) > 0$, for our parameters yields
\begin{align}
    \label{eqn:B3} 
    &\mathcal{B} = 2^{3|\ell| + 3r + 2} \left( \frac{1}{w_d^2 + 2w^2}\right)^{|\ell| + r +1} \frac{\Gamma\left( |\ell|+r + 1\right) \Gamma(|\ell| + p + 1)}{p! \Gamma(|\ell|  + 1)} \\
    &\times {}_2F_1 \left[ -p, |\ell|+r + 1 ; |\ell| + 1; \frac{4w^2}{w_d^2 + 2w^2}\right]. \nonumber
\end{align}
Where
\begin{equation}
    \label{s_constraint}
    \text{Re}\left(\frac{w_d^2 + 2w^2}{4w^2}\right) > 0,
\end{equation}
since $w_d$ and $w$ describe waist parameters which are strictly real valued, non-zero numbers, and
\begin{equation}
    \label{beta_constraint}
    \text{Re}(|\ell|+ r ) > -1,
\end{equation}
since $|\ell|$ and $r$ both range from $[0, \infty]$. Therefore, we can express the integral $\mathcal{N}^{\ell_i}_{p_i}$ as
\begin{align}
    \label{eqn:A6} 
    &\mathcal{N}^{\ell_i}_{p_i} = |C|^2 \sum_{r = 0}^{\infty} \frac{2^{2|\ell_i| + 5} \pi \Gamma^2(|\ell_i| + p_i + 1) \Gamma\left(|\ell_i| + r + 1\right)}{r! p_i!(p_i + |\ell_i|)! \Gamma^2(|\ell_i| + 1)}  w_d^{4r + 2|\ell_i|} (w_i w'_i)^{|\ell_i| + 1} \\
    &\times \left( \frac{1}{(w_d^2 + 2w_i^2)(w_d^2 + 2w_i'\,^2)}\right)^{|\ell_i| + r + 1}{}_2F_1 \left[ -p_i, |\ell_i| + r + 1 ; |\ell_i| + 1; \frac{4w_i^2}{w_d^2 + 2w_i^2}\right]\nonumber \nonumber \\ \nonumber &\times {}_2F_1 \left[ -p_i, |\ell_i| + r + 1 ; |\ell_i| + 1; \frac{4w'_i\,^2}{w_d^2 + 2w'_i\,^2}\right].
\end{align}  

\subsection{Solving the stimulated LG overlap integral}
\label{sec:LGInt2}
In this section, we analytically solve the overlap integrals in the stimulated term in Eq. (6) in the main text. Thus, solving the integral
\begin{align}
    \label{eqn:C1}
    &\mathcal{M}^{\ell_i, \ell_{se}}_{p_i, p_{se}} = C \int \!\!\! \int d^2\mathbf{q}_i d^2\mathbf{q}_{se} \tilde{W}_d(\mathbf{q}_i + \mathbf{q}_{se}) \tilde{LG}_{p_i}^{\ell_i} (\mathbf{q}_i) \tilde{LG}_{p_{se}}^{\ell_{se}} (\mathbf{q}_{se}) \\
    &= \int_{q_i = 0}^\infty dq_i \int_{\phi_i = 0}^{2\pi} q_i d\phi_i \int_{q_{se} = 0}^\infty dq_{se} \int_{\phi_{se} = 0}^{2\pi} q_{se} d\phi_{se} \frac{w_d}{\sqrt{2\pi}} e^{-w_d^2|\mathbf{q}_i + \mathbf{q}_{se}|^2/4} \sqrt{\frac{w_i^2 p_i!}{2\pi(p_i + |\ell_i|)!}} \nonumber \\
    &\times \sqrt{\frac{w_{se}^2 p_{se}!}{2\pi(p_{se} + |\ell_{se}|)!}} \left( \frac{w_i q_i}{\sqrt{2}} \right)^{|\ell_i|} \left( \frac{w_{se} q_{se}}{\sqrt{2}} \right)^{|\ell_{se}|} e^{-w_i^2 q_i^2 / 4} e^{-w_{se}^2 q_{se}^2 / 4} (-1)^{p_i + p_{se}} L_{p_i}^{|\ell_i|}\left(\frac{w_i^2 q_i^2}{2}\right)  \nonumber \\ \nonumber &\times L_{p_{se}}^{|\ell_{se}|}\left(\frac{w_{se}^2 q_{se}^2}{2}\right) e^{i\ell_i(\phi_i + \pi/2)} e^{i\ell_{se}(\phi_{se} + \pi/2)}.
\end{align}
This integral closely matches the integral solved earlier in Eq. (\ref{eqn:A2}) with a few minor changes: the kernel $\exp\{-w_d^2|\mathbf{q}_i - \mathbf{q'}_i|^2/8\}$ is replaced by the pump function $(w_d/\sqrt{2})\exp\{-w_d^2|\mathbf{q}_i+\mathbf{q}_{se}|^2/4\}$, the indices of the LG modes are not identical, and the modulus squared of the coefficients for the nonlinear process $|C|^2$ is replaced by $C$. Despite these changes, the same overall approach can be used to solve this integral. Solving the azimuthal integrals yields $\delta_{\ell', \ell_i}\delta_{\ell', -\ell_{se}}$ which forces the azimuthal phase-conjugation effect as the azimuthal index of the idler is constrained to be the negation of the azimuthal index of the seed photon $\ell_i = -\ell_{se}$. Following through with the rest of the calculation performed in Section (\ref{sec:LGInt1}), we can express the integral as
\begin{align}
    \label{eqn:C2} 
    &\mathcal{M}^{\ell_i, \ell_{se}}_{p_i, p_{se}} = \delta_{\ell_i,-\ell_{se}} C\sum_{r = 0}^{\infty} \frac{2^{|\ell_i| + 2} \Gamma(|\ell_i| + p_i + 1) \Gamma(|\ell_i| + p_{se} + 1) \Gamma\left(|\ell_i| + r + 1\right)}{r! \Gamma^2(|\ell_i| + 1)} \\
    &\times \sqrt{\frac{2\pi}{p_i! p_{se}!(p_i + |\ell_i|)! (p_{se} + |\ell_i|)!}} w_d^{4r + 2|\ell_i| + 1} (w_i w_{se})^{|\ell_i| + 1} (-1)^{p_i + p_{se} + |\ell_i|} \nonumber \\
    &\times \left( \frac{1}{(w_d^2 + w_i^2)(w_d^2 + w_{se}^2)}\right)^{|\ell_i| + r + 1} {}_2F_1 \left[ -p_i, |\ell_i| + r + 1 ; |\ell_i| + 1; \frac{2w_i^2}{w_d^2 + w_i^2}\right] \nonumber \\
    &\times {}_2F_1 \left[ -p_{se}, |\ell_i| + r + 1 ; |\ell_i| + 1; \frac{2w_{se}^2}{w_d^2 + w_{se}^2}\right] \nonumber.
\end{align}
\subsection{Measuring a single output mode from SPDC}
\label{sec:SPDCidler}    
\begin{figure}
    \centering
    \includegraphics[width=0.25\linewidth]{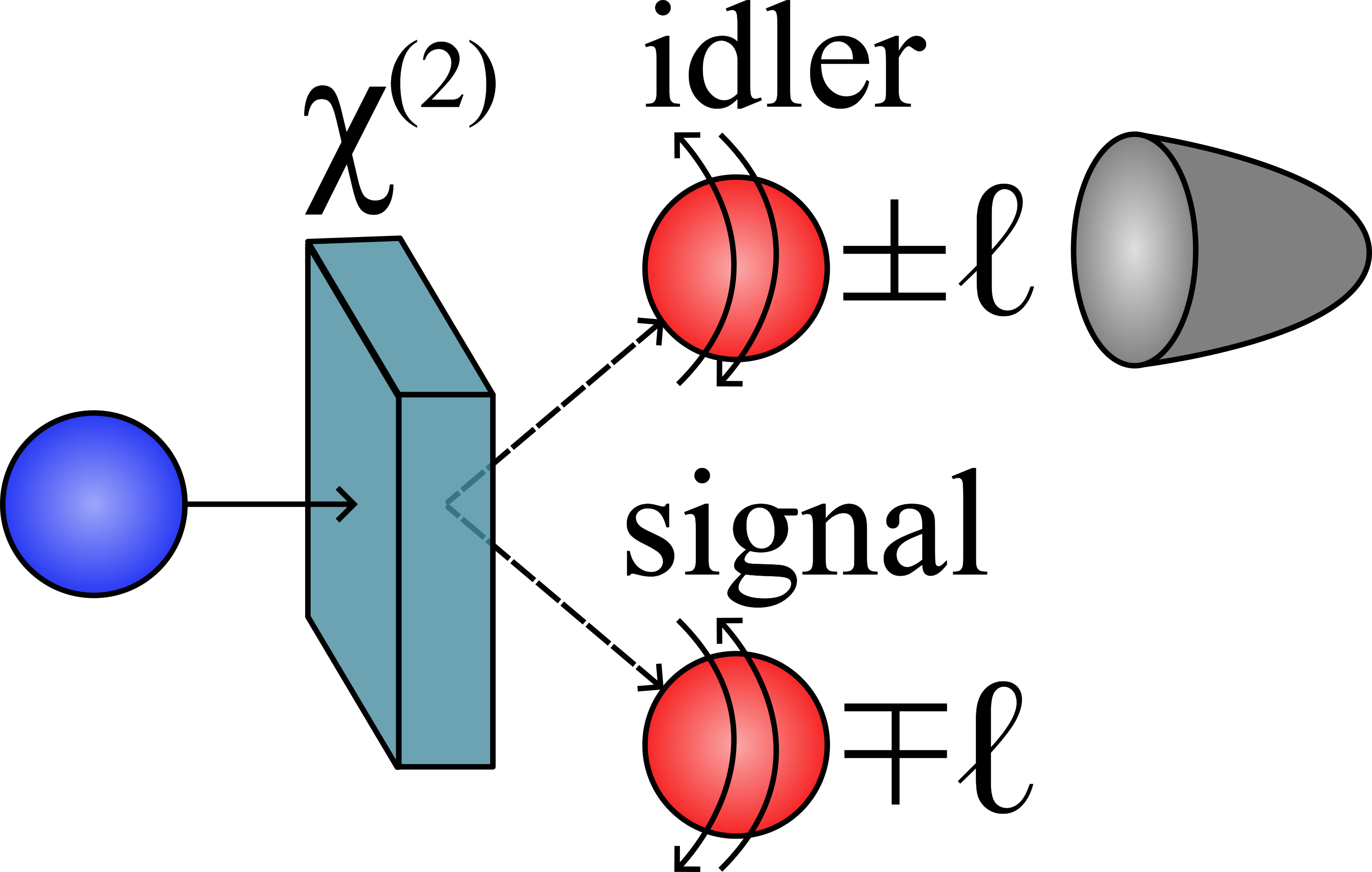}
    \caption{Schematic of the SPDC process in which the average number of photons in the idler mode
and in a specificed LG mode are measured.}
    \label{fig:SPDCidler}
\end{figure}

A term that appears multiple times in this study is $\mathcal{N}^\ell_p$, which corresponds to the triple overlap integral in Eq. (\ref{eqn:A1}). This term is distinct from the spiral spectrum and arises from a SPDC process in which only one of the output modes is measured. In this section, we derive how this term appears in SPDC when only measuring the idler mode, which gives the term more context for when it appears in other sections of the study. The process is modeled in Figure (\ref{fig:SPDCidler}).
The mean photon number associated with measuring down-converted idler photons in mode $\tilde{LG}^{\ell_i}_{p_i}$ can be found by calculating the overlap $[N^{\ell_i}_{p_i}]_{\text{SPDC}} = \langle \psi_{\text{SPDC}}|\hat{n}(\mathbf{q}_i, (\ell_i, p_i)) | \psi_{\text{SPDC}}\rangle$ where the biphoton SPDC state in the monochromatic, paraxial, and thin-crystal approximations is given by \cite{Riberiro:1999}
\begin{equation}
    \label{eqn:SPDCstate3}
    |\psi_{\text{SPDC}}\rangle  = |0\rangle + C \int \!\!\! \int d^2 \mathbf{q}_i d^2 \mathbf{q}_s \tilde{W}_d(\mathbf{q}_i + \mathbf{q}_s)|1;\mathbf{q}_i\rangle |1; \mathbf{q}_s \rangle.
\end{equation}
Due to the orthogonality of the Fock states: $\langle n; \mathbf{q}_1| m; \mathbf{q}_2\rangle = \delta(n - m)\delta^2(\mathbf{q}_1 - \mathbf{q}_2)$ the first three terms in the binomial expansion of the mean photon number vanish. Thus, the remaining term in the mean photon number is given by
\begin{align}
    \label{eqn:P_idler1}
    &[N^{\ell_i}_{p_i}]_{\text{SPDC}} = |C|^2 \int \!\! \dots \!\! \int d^2 \mathbf{q}_{i} d^2 \mathbf{q'}_{i} d^2 \mathbf{q''}_{i} d^2 \mathbf{q'''}_{i} d^2 \mathbf{q}_{s} d^2 \mathbf{q'}_{s} \tilde{W}_d(\mathbf{q}_i + \mathbf{q}_s) \tilde{W}^*_d(\mathbf{q'}_i + \mathbf{q'}_s) \\
    &\times \nonumber \left[ \tilde{LG}^{\ell_i}_{p_i}(\mathbf{q''}_i)\right]^* \tilde{LG}^{\ell_i}_{p_i}(\mathbf{q'''}_i) \langle 1; \mathbf{q'}_i| \hat{a}^\dagger(\mathbf{q''}_i) \hat{a}(\mathbf{q'''}_i)|1;\mathbf{q}_i\rangle \langle 1; \mathbf{q'}_s|1;\mathbf{q_s}\rangle.
\end{align}
The orthogonality of the annihilation and creator operators imposes $\delta^2(\mathbf{q'}_i - \mathbf{q''}_i)$ and $\delta^2(\mathbf{q'''}_i - \mathbf{q}_i)$ which collapses the integrals over $\mathbf{q''}_i$ and $\mathbf{q'''}_i$. Additionally, the orthogonality of the Fock states imposes the relation $\langle 1; \mathbf{q'}_s|1;\mathbf{q_s}\rangle = \delta^2(\mathbf{q'}_s - \mathbf{q}_s)$ The resulting mean photon number is then 
\begin{align}
    \label{eqn:P_idler2}
    &[N^{\ell_i}_{p_i}]_{\text{SPDC}} = |C|^2 \int \!\!\! \int \!\!\! \int d^2 \mathbf{q}_{i} d^2 \mathbf{q'}_{i} d^2 \mathbf{q}_{s} \tilde{W}_d(\mathbf{q}_i + \mathbf{q}_s) \tilde{W}^*_d(\mathbf{q'}_i + \mathbf{q}_s) \tilde{LG}^{\ell_i}_{p_i}(\mathbf{q}_i) \left[ \tilde{LG}^{\ell_i}_{p_i}(\mathbf{q'}_i)\right]^*,
\end{align}
which directly matches the form of the double integrals found in the first term in Eq. (6) in the main text and are analytically solved in Eq. (\ref{eqn:A6}).

\subsection{Spiral spectrum in SPDC}
\label{sec:SPDC}
The spiral spectrum in SPDC refers to the coincidence amplitudes associated with measuring down-converted idler-signal photon pairs in SPDC, $\mathcal{M}^{\ell_i, \ell_s}_{p_i, p_s}$\cite{Miatto:2011}. In the low-gain regime of SPDC, where the biphoton state contains exactly one signal-idler pair, the mean photon number of the joint number operator in specified Laguerre-Gaussian modes is equal to the detection probability for those modes. This is because the mean photon number reduces to the squared modulus of the projection amplitude of the biphoton state onto those modes, which directly yields the probability: $[N^{\ell_i, \ell_s}_{p_i, p_s}]_{\text{SPDC}} = [P^{\ell_i, \ell_s}_{p_i, p_s}]_{\text{SPDC}} = |\mathcal{M}^{\ell_i, \ell_s}_{p_i, p_s}|^2$. The mean photon number associated with measuring down-converted idler-signal photon pairs in specified LG modes can be found by calculating the overlap $[N^{\ell_i, \ell_s}_{p_i, p_s}]_{\text{SPDC}} = \langle \psi_{\text{SPDC}}|\hat{n}(\mathbf{q}_i, (\ell_i, p_i)) \hat{n}(\mathbf{q}_s, (\ell_s, p_s))| \psi_{\text{SPDC}}\rangle$ where the biphoton SPDC state in the monochromatic, paraxial, and thin-crystal approximations is given by \cite{Riberiro:1999}
\begin{equation}
    \label{eqn:SPDCstate2}
    |\psi_{\text{SPDC}}\rangle  = |0\rangle + C \int \!\!\! \int d^2 \mathbf{q}_i d^2 \mathbf{q}_s \tilde{W}_d(\mathbf{q}_i + \mathbf{q}_s)|1;\mathbf{q}_i\rangle |1; \mathbf{q}_s \rangle.
\end{equation}
Due to the orthogonality of the Fock states: $\langle n; \mathbf{q}_1| m; \mathbf{q}_2\rangle = \delta(n - m)\delta^2(\mathbf{q}_1 - \mathbf{q}_2)$ the first three terms in the binomial expansion of the mean photon number vanish. Thus, the remaining term in the mean photon number is given by
\begin{align}
    \label{eqn:P_SPDC1}
    &[N^{\ell_i, \ell_s}_{p_i, p_s}]_{\text{SPDC}} = CC^* \int \!\! \dots \!\! \int d^2 \mathbf{q}_{i} d^2 \mathbf{q'}_{i} d^2 \mathbf{q''}_{i} d^2 \mathbf{q'''}_{i} d^2 \mathbf{q}_{s} d^2 \mathbf{q'}_{s} d^2 \mathbf{q''}_{s} d^2 \mathbf{q'''}_{s} \tilde{W}_d(\mathbf{q}_i + \mathbf{q}_s)  \\
    &\times \nonumber \tilde{W}^*_d(\mathbf{q'}_i + \mathbf{q'}_s) \left[ \tilde{LG}^{\ell_i}_{p_i}(\mathbf{q''}_i)\right]^* \tilde{LG}^{\ell_i}_{p_i}(\mathbf{q'''}_i) \left[ \tilde{LG}^{\ell_s}_{p_s}(\mathbf{q''}_s)\right]^* \tilde{LG}^{\ell_s}_{p_s}(\mathbf{q'''}_s) \\ \nonumber &\times\langle 1; \mathbf{q'}_i| \hat{a}^\dagger(\mathbf{q''}_i) \hat{a}(\mathbf{q'''}_i)|1;\mathbf{q}_i\rangle \langle 1; \mathbf{q'}_s| \hat{a}^\dagger(\mathbf{q''}_s) \hat{a}(\mathbf{q'''}_s)|1;\mathbf{q_s}\rangle.
\end{align}
The action of the ladder operators on the Fock states and the orthogonality of the Fock states imposes $\delta^2(\mathbf{q'}_i - \mathbf{q''}_i)$, $\delta^2(\mathbf{q'''}_i - \mathbf{q}_i)$, $\delta^2(\mathbf{q'}_s - \mathbf{q''}_s)$, and $\delta^2(\mathbf{q'''}_s - \mathbf{q}_s)$ which collapses the integrals over $\mathbf{q''}_i$, $\mathbf{q'''}_i$, $\mathbf{q''}_s$, and $\mathbf{q'''}_s$. The resulting mean photon number is then 
\begin{align}
    \label{eqn:P_SPDC2}
    &[N^{\ell_i, \ell_s}_{p_i, p_s}]_{\text{SPDC}} = C \int \!\!\! \int d^2 \mathbf{q}_{i} d^2 \mathbf{q}_{s} \tilde{W}_d(\mathbf{q}_i + \mathbf{q}_s) \tilde{LG}^{\ell_i}_{p_i}(\mathbf{q}_i) \tilde{LG}^{\ell_s}_{p_s}(\mathbf{q}_s) C^* \!\!\! \int \!\!\! \int d^2 \mathbf{q'}_{i} d^2 \mathbf{q'}_{s} \tilde{W}^*_d(\mathbf{q'}_i + \mathbf{q'}_s) \\
    &\times \nonumber \left[\tilde{LG}^{\ell_i}_{p_i}(\mathbf{q'}_i)\right]^* \left[\tilde{LG}^{\ell_s}_{p_s}(\mathbf{q'}_s)\right]^* .
\end{align}
Since $[N^{\ell_i, \ell_s}_{p_i, p_s}]_{\text{SPDC}} = \mathcal{M}^{\ell_i, \ell_s}_{p_i, p_s} [\mathcal{M}^{\ell_i, \ell_s}_{p_i, p_s}]^*$ the coincidence amplitudes are then
\begin{align}
    \label{eqn:C_SPDC1}
    &\mathcal{M}^{\ell_i, \ell_s}_{p_i, p_s} = C \int \!\!\! \int d^2 \mathbf{q}_{i} d^2 \mathbf{q}_{s} \tilde{W}_d(\mathbf{q}_i + \mathbf{q}_s) \tilde{LG}^{\ell_i}_{p_i}(\mathbf{q}_i) \tilde{LG}^{\ell_s}_{p_s}(\mathbf{q}_s),
\end{align}
which directly matches the form of the double integrals found in the second term in Eq. (6) in the main text and are analytically solved in Eq. (\ref{eqn:C2}).

\subsection{Exploring the effects of differing beam waist ratios on the CPDC mean photon number distribution}
\label{sec:beamwaist}
\begin{figure}
    \centering
    \includegraphics[width=0.9\linewidth]{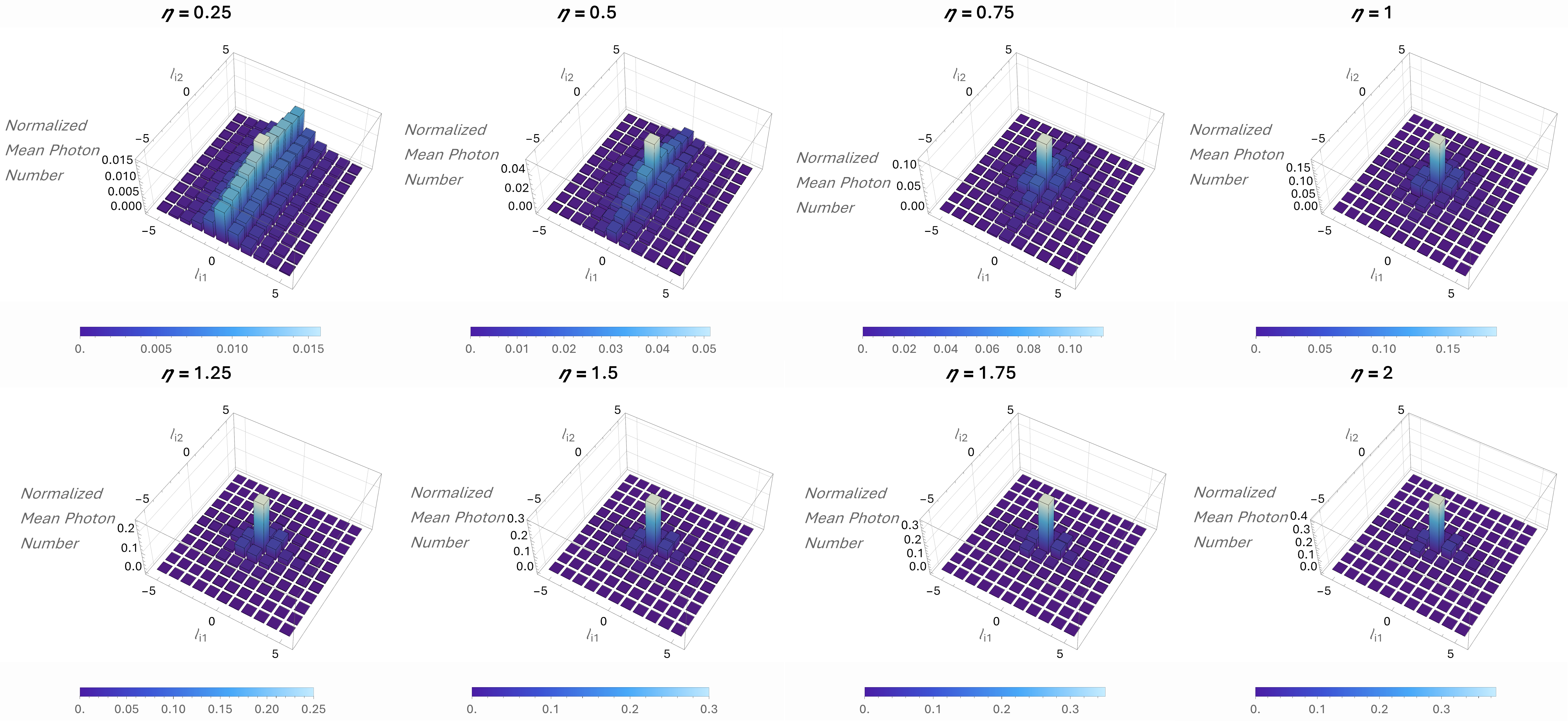}
    \caption{The simulated, joint normalized ($\sum_{\ell_{i1}, \ell_{i2}=-\infty}^\infty [N^{\ell_{i1}, \ell_{i2}}_{p_{i1}, p_{i2}}]_{\text{CPDC}} = 1$) mean photon number distribution $[N^{\ell_{i1}, \ell_{i2}}_{p_{i1}, p_{i2}}]_{\text{CPDC}}$ for $w_{d2} = w_{s1}$ and differing beam waist ratios, $\eta \in $ $[0.25, 0.5, 0.75, 1, 1.25, 1.5, 1.75, 2]$. The enhanced diagonal along $\ell_{i1}=\ell_{i2}$ is more prominent for values of $\eta > 1$. For values of $\eta < 1$, the distribution becomes more broad which obscures the diagonal enhancement. All distributions are plotted with radial indices $p_{i1}= p_{i2} = 0 $.}    
    \label{fig:beam_waists}
\end{figure}
In the main text, all mean photon number distributions were plotted assuming the detection beam waists are equal to the beam waist of the pump ($w_{d2} = w_{i2} = w'_{i2} = w_{s1} = w_{s2}$) for simplicity. However, previous studies have explored the effects of changing the ratios between the different beam waists on the coincidence amplitudes in SPDC and StimPDC \cite{Miatto:2011, Xu:2024}. Since the phase conjugation effect of interest occurs at the second nonlinear crystal, we will investigate the cases for which $w_{d2} = w_{s1}$ and all other detection beam waists are equal to the same multiple of the pump beam waist, $w_{d2} = \eta w_{i2} = \eta w'_{i2} = \eta w_{s2}$.

The mean photon number distribution as a function of $\ell_{i1}$ and $\ell_{i2}$ in the CPDC configuration is plotted in Fig. (\ref{fig:beam_waists}) for different detection beam waists. The presence of the enhanced diagonal for values of $\eta > 1$ indicates how the beam waist ratio at the second nonlinear crystal can be altered to enhance the phase conjugation effect. We also observe an enhanced horizontal along $\ell_{i2} = 0$ becoming increasingly more prominent as $\eta$ increases. This is because the weight $\mathcal{N}^{\ell_{i2}}_{p_{i2}}$ in the spontaneous term in $[N^{\ell_{i1}, \ell_{i2}}_{p_{i1}, p_{i2}}]_{\text{CPDC}}$  tightens quicker about $\ell_{i2} = 0$ with increasing $\eta$ than $\mathcal{M}^\ell_p$ does. In contrast, values of $\eta < 1$ obscure the enhanced diagonal as both the spontaneous and stimulated terms broaden which increases the relative weight to the off-diagonal terms, thus reducing the relative weight of the diagonal terms upon normalization.

\end{document}